\documentclass[sigplan,10pt, nonacm]{acmart}
\AtBeginDocument{%
  }

\usepackage{sections/utils}

\begin{document}


\newcommand{\sysname}{TaiChi\xspace}

\title{Prefill-Decode Aggregation or Disaggregation? Unifying Both for Goodput-Optimized LLM Serving}

\author{Chao Wang}
\authornote{Work done during their internship at Huawei Cloud.}
\affiliation{%
  \institution{The Chinese University of Hong Kong}
  \country{} 
}

\author{Pengfei Zuo}
\authornote{Corresponding author is Pengfei Zuo (pengfei.zuo@huawei.com).}
\affiliation{%
  \institution{Huawei Cloud}
  \country{} 
}

\author{Zhangyu Chen}
\affiliation{%
  \institution{Huawei Cloud}
  \country{} 
}

\author{Yunkai Liang}
\authornotemark[1]
\affiliation{%
  \institution{Sun Yat-sen University}
  \country{} 
}

\author{Zhou Yu}
\affiliation{%
  \institution{Huawei Cloud}
  \country{} 
}

\author{Ming-Chang Yang}
\affiliation{%
  \institution{The Chinese University of Hong Kong}
  \country{} 
}

\begin{abstract}

There is an ongoing debate on whether prefill-decode (PD) aggregation or disaggregation is the superior approach for serving large language models (LLMs). This debate has driven optimizations on both sides, each showcasing distinct advantages.
This paper presents a comprehensive comparison between PD aggregation and disaggregation, showing that each excels under different service-level objectives (SLOs): PD aggregation is optimal under tight time-to-first-token (TTFT) and relaxed time-per-output-token (TPOT), while PD disaggregation excels under strict TPOT and relaxed TTFT. However, under balanced TTFT and TPOT SLOs, neither approach can deliver optimal goodput.

Based on these insights, this paper proposes \sysname, an LLM serving system that unifies PD disaggregation and aggregation to achieve optimal goodput under any combination of TTFT and TPOT SLOs. 
\sysname leverages a unified disaggregation-aggregation architecture composed of differentiated-capability GPU instances: prefill-heavy instances (fast prefill but high-interference decode) and decode-heavy instances (low-interference decode but slow prefill). It exposes three configurable sliders to control the ratio between prefill-heavy and decode-heavy instances, and the chunk sizes for each. \sysname adapts to various SLO regimes by adjusting these sliders. When TTFT constraints are tight, \sysname can be tuned to resemble a PD aggregation configuration; when TPOT dominates, it adapts toward PD disaggregation. Crucially, under balanced SLOs, \sysname enables a hybrid mode that achieves superior goodput. The key innovation behind this hybrid mode is latency shifting: by selectively reallocating GPU resources from requests that meet TTFT or TPOT SLOs to those at risk of violation, \sysname maximizes the number of SLO-satisfied requests. This fine-grained, request-level latency shifting is orchestrated through two targeted scheduling mechanisms: flowing decode scheduling to control TPOTs and length-aware prefill scheduling to manage TTFTs, jointly optimizing request assignment.
Our extensive experimental results demonstrate that \sysname improves goodput by up to 77\% compared to state-of-the-art systems under balanced TTFT and TPOT SLOs.

\end{abstract}

\settopmatter{printfolios=true}
\maketitle


\vspace{-1em}

\section{Introduction}

Large language models (LLMs) have demonstrated unprecedented performance across various applications, such as personal assistants~\cite{PersonalAssistant}, translation~\cite{Translation1, Translation2}, document analysis~\cite{DocumentAnalysis}, chatbots~\cite{ChatGPT, GPT4}, and code generators~\cite{CodeGen1, CodeGen2}. 
Serving these LLM applications requires substantial and costly computational resources, particularly GPUs.
Consequently, optimizing the LLM serving cost has received significant attention in system research~\cite{vLLM, Orca, ChunkedPrefill, DistServe, Dejavu, SplitWise,gao2024cachedattention}.

Meeting service level objective (SLO) constraints is critical for LLM service providers to ensure application-level performance~\cite{DistServe,ChunkedPrefill}. 
In LLM serving, a user request is processed in two distinct phases, each with its own SLO constraint~\cite{DistServe}. The first phase, known as \textit{prefill}, involves preprocessing the user request and generating the first token with intensive computation. The latency of prefill is constrained by time-to-first-token (TTFT), reflecting system responsiveness~\cite{MetricSurvey, DistServe}. The second phase, known as \textit{decode}, outputs the subsequent tokens autoregressively (i.e., one token per iteration). The average time taken to output a token (except for the first token) is constrained by time-per-output-token (TPOT), which indicates the service speed users can experience~\cite{MetricSurvey, DistServe}.
The throughput of requests while meeting both SLO constraints is referred to as \textit{goodput}~\cite{MetricSurvey, DistServe}. Enhancing goodput with the same hardware resources lowers the cost per LLM request~\cite{DistServe}. However, application-level SLOs vary widely~\cite{ChunkedPrefill,DistServe,SOLA,Tempo2025}: some prioritize low TTFT with relaxed TPOT, others require tight TPOT with relaxed TTFT, while many demand a balanced trade-off between the two.

\begin{figure}[t]
    \centering
    \includegraphics[width=0.4\textwidth]{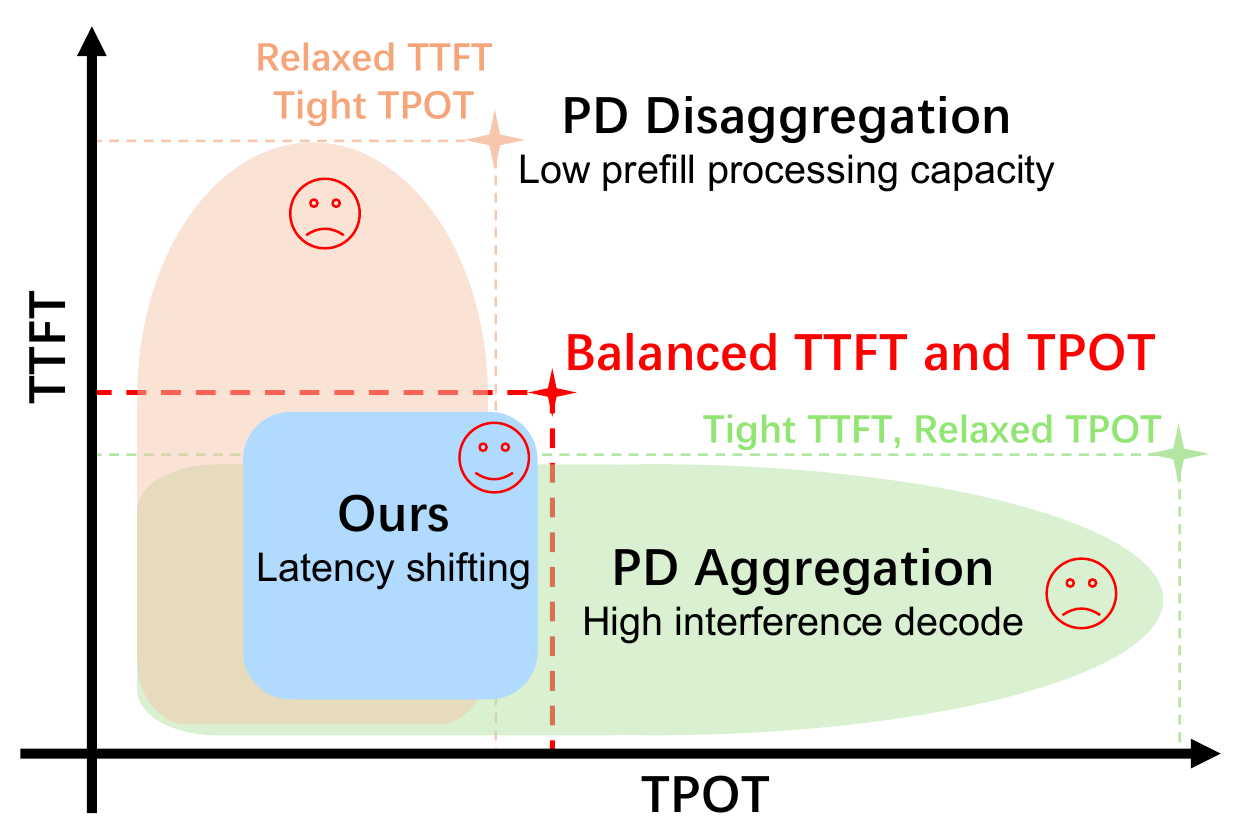}
    \caption{Distribution of requests' TTFT and TPOT under different scheduling approaches, using the same number of compute nodes and QPS. \textit{(PD aggregation performs best when TTFT constraints are tight and TPOT is relaxed, while PD disaggregation excels under tight TPOT and relaxed TTFT. However, under balanced SLO constraints, PD aggregation results in TPOT violations due to high-interference decode, while PD disaggregation leads to TTFT violations due to low prefill processing capacity. By proposing a hybrid-mode inference, we mitigate both issues and achieve better SLO attainment across different combinations of TTFT and TPOT.)}}
    \label{fig/introduction/category}
    \Description{None}
\end{figure}

Currently, an active debate is ongoing regarding whether PD aggregation~\cite{ChunkedPrefill} or disaggregation~\cite{DistServe} is the superior approach for serving LLMs. PD aggregation (like Orca~\cite{Orca} and Sarathi-Serve~\cite{ChunkedPrefill}) co-locates the prefill and decode phases of a request on the same hardware instance to achieve high resource utilization. In contrast, PD disaggregation (like Splitwise~\cite{SplitWise} and DistServe~\cite{DistServe}) physically separates prefill and decode onto different hardware instances. This approach eliminates interference between the two phases and allows for independent scaling of resources.
The debate drives optimizations on both sides, each showcasing the distinct advantages~\cite{ChunkedPrefill, DistServe, SplitWise}.

In this paper, we present a comprehensive comparison between PD aggregation and disaggregation, showing that each approach achieves optimal request goodput under different SLOs: PD aggregation is optimal under tight TTFT and relaxed TPOT, while PD disaggregation excels under strict TPOT and relaxed TTFT, as illustrated in Figure~\ref{fig/introduction/category}. PD aggregation achieves low TTFT by having all instances participate in the prefill phase, but suffers in TPOT due to interference between prefill and decode. In contrast, PD disaggregation improves TPOT by isolating prefill and decode on separate resources, but incurs higher TTFT since fewer instances handle prefill. 
However, under balanced TTFT and TPOT SLOs, neither approach can deliver optimal goodput. This is because PD aggregation tends to violate TPOT due to decode interference, while PD disaggregation often fails to meet TTFT, as only a subset of instances handle the prefill phase.

To end this debate about these two branches, we propose \sysname, an LLM serving system that unifies PD disaggregation and aggregation to achieve optimal goodput under any combination of TTFT
and TPOT SLOs. \sysname leverages a unified disaggregation-aggregation architecture composed of differentiated-capability GPU instances: P-heavy instances (fast prefill but high-interference decode) and D-heavy instances (low-interference decode but slow prefill). It exposes three configurable sliders to control the ratio between P-heavy and D-heavy instances, and the chunk sizes for each. \sysname adapts to various SLO regimes by adjusting these sliders. When TTFT constraints are tight, \sysname can be tuned to resemble a PD aggregation configuration; when TPOT dominates, it adapts toward PD disaggregation. Crucially, under balanced SLOs, \sysname enables a hybrid mode that achieves superior goodput. The key innovation behind this hybrid mode is to shift the latency (i.e., TTFT and TPOT) across the prefill and decode phases, as well as across requests.
This means that we strategically degrade the latency of requests already meeting SLO constraints, thereby reallocating overprovisioned GPU resources (i.e., GPU time) to prioritize SLO-violating requests through scheduling.
However, implementing this request-level latency degradation faces three main challenges: 

\textbf{1) Lack of Architecture Support for Latency Shifting.} Existing methods~\cite{DistServe,ChunkedPrefill} lack architectural support, offering no flexibility in scheduling to reallocate latency across requests. Both PD aggregation and disaggregation approaches rely on uniform GPU instance configurations dedicated either to prefill or decode, preventing differentiated treatment of individual requests. As a result, systems cannot selectively optimize or degrade requests based on their SLO urgency, limiting their ability to shift latency where it is most needed.

\textbf{2) Request-level TPOT Degradation Hindered by Batching and Output Length Uncertainty.} Degrading TPOT at the granularity of individual requests is challenging due to the constraints of batch processing and the unpredictability of output lengths. In batch decode, performance optimizations or degradations applied to one request inevitably affect all co-located requests, regardless of whether they benefit from or can tolerate such changes. This lack of isolation limits the system’s ability to selectively degrade TPOT. Furthermore, decode requests with shorter output lengths are more susceptible to prefill-decode interference (as discussed in \S~\ref{subsec/challenges}) and should avoid TPOT degradation. However, since the output length of a request is unknown in advance, the scheduler lacks the necessary information to make precise, per-request degradation decisions.

\textbf{3) Selective TTFT Degradation Constrained by Execution and Queuing Times.} Selectively degrading TTFT is complicated by the need to consider both execution and queuing times.
Long prefill requests inherently consume more execution time and are more likely to violate TTFT constraints, making them poor candidates for further degradation. Similarly, requests that have already spent considerable time in the queue are at higher risk of SLO violation and should also be protected. Therefore, determining which requests can safely tolerate TTFT degradation requires careful, context-aware scheduling.

Together, these factors hinder fine-grained, request-level latency control in existing LLM serving systems. To address them efficiently, \sysname introduces the following techniques:

\textbf{1) Hybrid-Mode Inference.}
To address Challenge 1, we introduce hybrid-mode inference to enable latency shifting. This approach uniquely combines the advantages of both PD aggregation and disaggregation. To achieve high resource efficiency, it allows all specialized instances—both P-heavy and D-heavy—to process mixed batches containing both prefill and decode tasks, thereby maximizing GPU utilization. Simultaneously, to provide fine-grained control, it adopts the flexibility of disaggregation, allowing the prefill and decode phases of a single request to be executed on different instance types. This decoupling enables strategic latency shifting—the ability to trade latency between the two phases or across different requests. For instance, a request's TTFT can be minimized by processing its prefill on a P-heavy instance, while its decode phase is handled by a D-heavy instance to ensure a low TPOT. This core capability underpins the advanced scheduling techniques that follow.

\textbf{2) Flowing Decode Scheduling.}
To address Challenge 2, we propose flowing decode scheduling, which selectively degrades TPOT by dynamically migrating decode requests between D-heavy and P-heavy instances, enabling fine-grained, per-request latency control without cross-request interference. All decode requests are initially assigned to D-heavy instances to prevent unrecognizable short-output requests from completing the decode phase on P-heavy instances and avoid premature TPOT violations. To prevent the degradation of a request's TPOT from impacting others in the same batch, we extract the selected request from its batch in the low-interference (D-heavy) instance and migrate it to a high-interference (P-heavy) instance, thus strategically degrading its TPOT. 
Since output lengths are unknown a priori, we employ a longest-first approach, which selects the request with the current longest output in the D-heavy instance for degradation, as it has the greatest remaining TPOT budget currently and can better absorb performance degradation.
Finally, to prevent over-degradation, we monitor the TPOT of migrated requests in real time. Once the TPOT approaches the SLO constraint, the request is flowed back to a D-heavy instance to preserve service quality.

\textbf{3) Length-Aware Prefill Scheduling.}
To address Challenge 3, we propose a length-aware prefill scheduling strategy, which selectively degrades TTFT by assigning short prefill requests to slower instances when doing so does not violate SLO constraints. The key idea is to exploit the lower urgency of short prefill requests by routing them to D-heavy instances, intentionally slowing their execution to free up P-heavy instances for more time-sensitive, long prefill requests. To determine whether a short request is degradable, we estimate its projected TTFT on each D-heavy instance by summing the expected queuing delay and degraded execution time. If this total remains within the TTFT SLO, the request is marked as degradable and scheduled accordingly. 

We have implemented \sysname on vLLM~\cite{vllm_project} and plan to open-source it in the near future. Experiments show \sysname improves goodput by up to 77\% over SOTA systems. It also reduces TTFT by up to $13.2\times$ and TPOT by up to $1.69\times$, relative to PD disaggregation and PD aggregation, respectively.

The main contributions of this paper are as follows:
\begin{enumerate}
    \item We identify a fundamental trade-off in existing systems between optimizing TTFT and TPOT, which limits overall goodput.
    \item We propose the \sysname, a unified LLM serving system that leverages a hybrid aggregation-disaggregation architecture and latency-shifting scheduling policies to resolve this trade-off.
    \item We demonstrate the advantages of \sysname through comprehensive experiments.
\end{enumerate}

\section{Background and Motivation}
\label{sec/background}

\subsection{LLM Inference}

\textbf{Transformer Architecture.} 
The popular LLMs such as GPT-4~\cite{GPT4} and LLaMA~\cite{Llama2} are built upon decoder-only transformer models, which are optimized for next-token prediction~\cite{ChunkedPrefill}. These models consist of a stack of identical layers, each including a self-attention mechanism and a feed-forward network (FFN). In each layer, the self-attention module computes contextualized token embeddings by attending over all previous tokens. This involves computing query (Q), key (K), and value (V) vectors and applying scaled dot-product attention. The resulting vector is passed through an FFN block to produce the output embedding for the next layer. Notably, the K and V vectors are cached during decode to avoid recomputation, forming the key-value (KV) cache used for efficient generation.

\textbf{Two-Phase Inference.} 
LLM inference proceeds in two distinct stages: the \textit{prefill} phase processes the full prompt in parallel to generate the first token, and the \textit{decode} phase generates subsequent tokens one by one. Prefill is compute-intensive, leveraging the parallelism of the transformer to fully utilize the GPU across the input sequence. In contrast, decode is memory-bound and sequential, as it processes one token at a time using cached key/value vectors of the previous tokens. This asymmetry leads to under-utilization of compute during decode. 

\textbf{Batching.} 
To improve GPU utilization, LLM serving systems batch multiple inference requests, particularly for the decode phase. Batching amortizes model loading costs and maximizes throughput, especially during decode, where token-wise generation is lightweight. However, batching heterogeneous requests introduces latency variability. Fixed-size request-level batching is simple but inefficient, as longer requests delay batch progress. To solve it, continuous batching~\cite{Orca} is proposed to allow requests to enter and exit batches dynamically, which improves GPU occupancy.

\textbf{Performance Metrics.}
Latency service-level objectives (SLOs) quantify user-perceived performance by specifying bounds on time-to-first-token (\textbf{TTFT}) and time-per-output-token (\textbf{TPOT}). TTFT measures the latency from request arrival to the first token, primarily determined by prefill time, while TPOT reflects the average per-token latency during decode. Satisfying these SLOs is essential for interactive responsiveness and smooth token streaming~\cite{MetricSurvey}.
\textbf{Goodput} denotes the maximum request rate that can be sustained while meeting SLO targets~\cite{DistServe}. This metric is critical as it influences serving costs: increasing goodput on fixed hardware lowers the cost per query.

\begin{table}[t]
  \centering
  \caption{A comparison of different scheduling approaches.}
  \label{tab/policy-comparison}
  \small
  \begin{tabular}{@{}ccc@{}}
    \toprule
    \textbf{Scheduling Policy} & \textbf{Batch} & \textbf{Request} \\
    \midrule
    PD Aggregation~\cite{Orca,ChunkedPrefill}     & Aggregated      & Aggregated      \\
    PD Disaggregation~\cite{DistServe,SplitWise}     & Disaggregated   & Disaggregated   \\
    \textbf{Hybrid Mode} & \textbf{Aggregated}    & \textbf{Disaggregated} \\
    \bottomrule
  \end{tabular}
\end{table}

\subsection{Scheduling Policies for LLM Serving}
\label{subsec/background/scheduling_policies}

The scheduler determines how requests are batched and assigned across the LLM serving instances. Modern policies can be grouped into two main categories: \textit{prefill-decode (PD) aggregation} and \textit{PD disaggregation}, depending on whether the two phases of the requests share the same hardware instance.

\textbf{PD Aggregation.} 
Most existing systems, including Orca~\cite{Orca}, and Sarathi-Serve~\cite{ChunkedPrefill}, colocate prefill and decode on the same GPU instance for high resource utilization. Orca improves utilization with iteration-level batching, where requests may join or leave the batch after each iteration. However, a new prefill request can dominate a full iteration, potentially delaying ongoing decode tasks for an extended period. To address this, Sarathi-Serve proposes \textit{chunked prefill}, which divides prefill into smaller chunks that are piggybacked in decode batches as additional computation. This piggybacking approach improves compute resource utilization during decode~\cite{ChunkedPrefill}.

\textbf{PD Disaggregation.} 
Recent systems such as Splitwise~\cite{SplitWise} and DistServe~\cite{DistServe} physically separate prefill and decode across different hardware instances, eliminating prefill-decode interference and enabling independent scaling. After the first token is computed, the KV-cache is transferred from the prefill to the decode instance via high-speed interconnects. Advances in interconnects (e.g., inter-GPU NVLINK at 600 GB/s, inter-node InfiniBand at 800 Gbps) and memory-efficient attention mechanisms (e.g., GQA~\cite{GQA}, MLA~\cite{MLA}) have made this transfer overhead negligible~\cite{DistServe}. As a result, PD disaggregation offers greater scheduling flexibility and higher goodput under strict TPOT constraints.

As outlined in Table~\ref{tab/policy-comparison}, these scheduling policies can be analyzed in two key dimensions: batch handling and request handling. Batch handling determines whether a batch mixes prefill and decode computations (aggregated) or is specialized for a single phase (disaggregated). An aggregated batch can improve GPU utilization, while a disaggregated batch eliminates interference between phases. The request handling defines whether a request's prefill and decode phases are treated as a single scheduling unit (aggregated) or can be separated across different GPU instances (disaggregated). An aggregated request is simpler to schedule, whereas a disaggregated one provides greater scheduling flexibility.

\subsection{Dilemma of Existing Methods}
\label{sec/motivation}

The improvement of goodput in LLM serving systems is constrained by two SLO constraints: TTFT and TPOT. Optimizing for only one SLO constraint may cause the other SLO constraint to become the bottleneck for improving goodput. Our investigation of PD aggregation and disaggregation reveals the following dilemma:

\textbf{Observation 1:}
\textit{
PD aggregation performs best under tight TTFT and relaxed TPOT constraints, while PD disaggregation is more effective under tight TPOT and relaxed TTFT. However, when TTFT and TPOT constraints are balanced, both approaches struggle to meet SLOs effectively.
}

\begin{figure}[t]
    \centering
    \begin{subfigure}[b]{0.23\textwidth}
        \includegraphics[width=\textwidth]{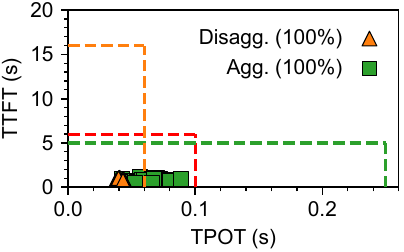}
        \caption{Low QPS level (QPS=6)}
        \label{fig/motivation/scatter_a}
    \end{subfigure}
    \hfill
    \begin{subfigure}[b]{0.23\textwidth}
        \includegraphics[width=\textwidth]{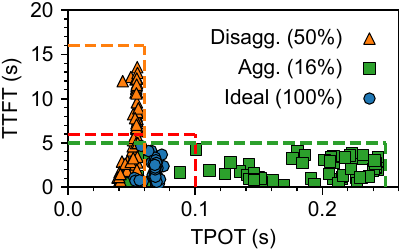}
        \caption{High QPS level (QPS=12)}
        \label{fig/motivation/scatter_b}
    \end{subfigure}
    \caption{TTFT and TPOT request distributions for different approaches across varying QPS levels, under multiple SLO constraints (orange for relaxed TTFT and tight TPOT, green for tight TTFT and relaxed TPOT, red for balanced SLOs). The attainment rates (\%) under balanced SLOs are in parentheses.}
    \label{fig/motivation/scatter}
    \Description{None}
\end{figure}

\begin{table}[t]
  \centering
  \caption{SLO attainment rates of different scheduling approaches under varying TTFT and TPOT SLOs (QPS=12).}
  \label{tab/slo-attainment-comparison}
  \footnotesize
  \begin{tabular}{@{}ccc@{}}
    \toprule
    \textbf{TTFT \& TPOT SLOs} & \textbf{PD Aggregation} & \textbf{PD Disaggregation} \\
    \midrule
    \makecell[c]{Relaxed TTFT \& Tight TPOT\\(16s, 60ms)}     &   7\%    &  98\%     \\
    \midrule
    \makecell[c]{Tight TTFT \& Relaxed TPOT\\(5s, 250ms)}     &  97\%    &  42\%     \\
    \midrule
    \makecell[c]{Balanced TTFT \& TPOT\\(6s, 100ms)}          &   16\%   &  50\%     \\
    \bottomrule
  \end{tabular}
\end{table}

To investigate their performance characteristics, we conduct experiments using Vidur\footnote{Vidur is an LLM inference simulator~\cite{Vidur} that emulates kernel latency with high accuracy (<3\% error), providing rich performance insight. All experiments in this section are performed on a 4-node, 8-GPU A100-DGX cluster, deploying the Llama-2-70B model~\cite{Llama2} with 4-way tensor parallelism (TP4). The Arxiv summarization dataset~\cite{Arxiv} is used, limiting requests to under 4096 tokens to fit the model’s context window.}.
Figure~\ref{fig/motivation/scatter} presents the TTFT and TPOT distributions for individual requests under both PD aggregation and disaggregation at different query per second (QPS) levels. 
As the load increases (increasing QPS from 6 to 12), PD disaggregation exhibits a significant elongation of TTFT, whereas PD aggregation shows a considerable increase in TPOT.
To quantify this performance degradation, we evaluate the SLO attainment rate for both schemes under high load (QPS=12) against three distinct sets of SLO constraints, as detailed in Table~\ref{tab/slo-attainment-comparison}.
Under a relaxed TTFT (e.g., 16 s) and a tight TPOT (e.g., 60 ms) constraint, PD disaggregation performs admirably, achieving a 98\% SLO attainment rate, while PD aggregation only reaches 7\%.
Conversely, with a tight TTFT (e.g., 5s) and a relaxed TPOT (e.g., 250ms), PD aggregation achieves a high attainment rate of 97\%, whereas PD disaggregation only attains 42\%.
These findings indicate that both PD disaggregation and aggregation excel only when one SLO metric is strictly constrained while the other is relaxed.
However, under moderately balanced dual SLO constraints (TTFT=6s, TPOT=100ms), the SLO attainment rates for PD disaggregation and aggregation drop to 50\% and 16\%, respectively. This demonstrates that neither approach can effectively satisfy balanced SLO requirements.
Considering that both TTFT and TPOT are crucial to user experience, impacting perceived responsiveness and fluency, it is imperative to conduct further analysis and optimization to improve the goodput in scenarios with balanced SLOs.

In the rest of this section, we further investigate the underlying causes of PD aggregation's struggle to meet TPOT constraints (\S~\ref{subsec/motivation/PD_aggregation}) and PD disaggregation's challenges in satisfying the TTFT constraints (\S~\ref{subsec/motivation/PD_disaggregation}).


\subsubsection{TPOT Bottleneck of PD Aggregation}
\label{subsec/motivation/PD_aggregation}

PD aggregation demonstrates excellent TTFT but suffers from high TPOT. To investigate the underlying causes of this TPOT degradation, we conduct a series of diagnostic experiments and obtain \textit{Observation 2}.

\textbf{Observation 2:} 
\textit{
The high TPOT in PD aggregation arises from prefill interference due to computation-bound linear operations, with a strong linear relationship between interference intensity and TPOT.
}

\begin{figure}[t!]
   \begin{minipage}{0.48\linewidth}
     \centering
     \includegraphics[width=0.98\linewidth]{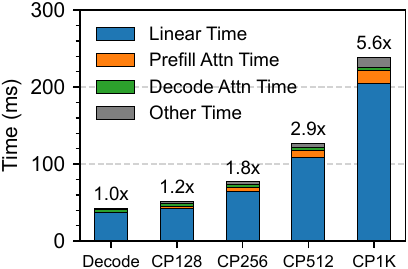}
    \caption{\label{fig/motivation/iteration_breakdown} Breakdown of batch execution time with varying chunk sizes (batch size = 16). \textit{CPxxx} refers to Chunked Prefill with a chunk size of xxx.}
    \Description{None}
   \end{minipage}
   \hfill
   \begin{minipage}{0.48\linewidth}
     \centering
     \includegraphics[width=0.98\linewidth]{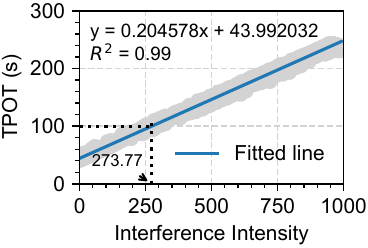}
    \caption{\label{fig/motivation/tpot_quantitative} Scatter plot of the requests' TPOT and their suffered interference intensity (CP1024), with a fitted line (R² = 0.99, strong correlation).}
    \Description{None}
   \end{minipage}
\end{figure}

Figure~\ref{fig/motivation/iteration_breakdown} shows the temporal breakdown of batch execution time with varying chunk sizes. As the chunk size increases, the total execution time increases. This is because larger chunk sizes introduce more prefill tokens into the batch, which leads to increased time spent on linear operations (i.e., matrix multiplications) associated with prefill.

To better understand the relationship between TPOT and prefill-decode interference, we perform a quantitative analysis. Before that, we define \textit{interference intensity} as the ratio of total prefill tokens computed during a decode request to its output length, measured in \textit{prefill tokens per output token}. For example, if a decode request generates 100 output tokens but 50,000 prefill token computations occur concurrently, its interference intensity is 500 (50,000/100) prefill tokens per output token.
Figure 4 reveals a strong linear correlation between TPOT and interference intensity, as evidenced by the fitted line's R-squared value of 0.99.
The slope of the fitted line's equation represents the increase in TPOT per additional token of interference intensity (here 0.2 ms). The intercept indicates the decode time in the absence of interference (here 44 ms). Furthermore, Figure 4 suggests that the key to controlling TPOT lies in regulating interference intensity. For instance, if we aim to keep TPOT below 100 ms, we must limit interference intensity to fewer than 273.77 prefill tokens per output token.

Although smaller chunk sizes reduce interference by limiting the maximum interference per decode token (or batch), they are not always optimal. As shown in Figure~\ref{fig/motivation/pd_aggregation_scatter}, chunk sizes below 1024 (e.g., 128, 256, 512) constrain TPOT but result in prohibitively high TTFT (analyzed later in \S~\ref{subsec/motivation/PD_disaggregation}), making the system unsustainable for the workload.
Thus, the optimal configuration should adopt the smallest chunk size that still satisfies the TTFT constraint.

\subsubsection{TTFT Bottleneck of PD Disaggregation}
\label{subsec/motivation/PD_disaggregation}

While PD disaggregation achieves high TPOT, it is easy to violate the TTFT SLO constraint. Through systematic experiments, we analyze the root cause of elevated TTFT in PD disaggregation and have \textit{Observation 3}.

\textbf{Observation 3}: 
\textit{
The high TTFT in PD disaggregation stems from request queuing, which occurs because PD disaggregation offers lower prefill processing capacity compared to PD aggregation.
}

Figure 6 shows the performance distribution of TTFT and TPOT in PD disaggregation with different PD ratios, compared with those of PD aggregation (CP1024). Experimental results demonstrate that when the PD ratio is adjusted from 4:4 to 7:1, TTFT exhibits a non-monotonic trend of initial decrease followed by an increase. In all configurations, PD disaggregation results in higher TTFT than PD aggregation.

Figure 7 further breaks down the p90 TTFT, revealing that queuing time (including both prefill and decode queues; note that decode queuing time is included in TTFT, as users experience it only once, following the same measurement as vLLM~\cite{vllm_project}) dominates TTFT in PD disaggregation.
Significant queuing times indicate the system has surpassed its processing capacity: high prefill queue times result from insufficient prefill processing resources, while decode queue latency reflects inadequate memory in decode instances.
Notably, PD ratios that cause queuing in the decode queue should be excluded, since decode requests generally take much longer to execute than prefill requests.

\begin{figure}[t!]
   \begin{minipage}{0.48\linewidth}
     \centering
     \includegraphics[width=0.98\textwidth]{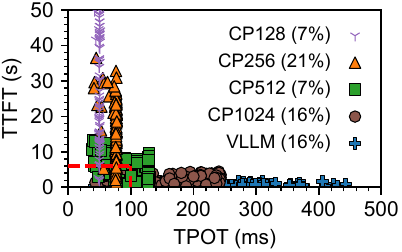}
     \caption{\label{fig/motivation/pd_aggregation_scatter} Latency distribution under varying configurations of PD aggregation (QPS=12).}
     \Description{None}
   \end{minipage}
   \hfill
   \begin{minipage}{0.48\linewidth}
     \centering
     \includegraphics[width=0.98\textwidth]{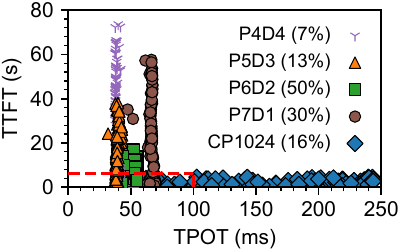}
     \caption{\label{fig/motivation/pd_disaggregation_scatter} Latency distribution under varying configurations of PD disaggregation (QPS=12).}
     \Description{None}
   \end{minipage}
\end{figure}

\begin{figure}[t!]
   \begin{minipage}[t]{0.48\linewidth}
     \centering
     \includegraphics[width=0.98\textwidth]{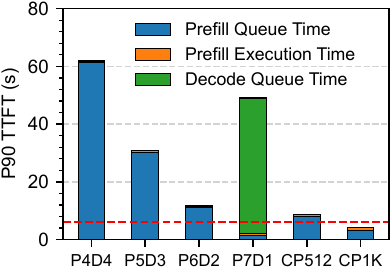}
     \caption{\label{fig/motivation/queue} Breaking down the P90 tail TTFT of PD disaggregation (PxDy) and PD aggregation (CPxxx).}
     \Description{None}
   \end{minipage}
   \hfill
   \begin{minipage}[t]{0.48\linewidth}
     \centering
     \includegraphics[width=0.98\textwidth]{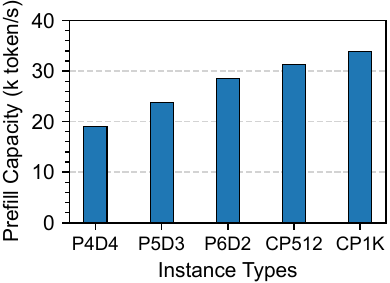}
     \caption{\label{fig/motivation/prefill_capacity} Prefill processing capacity of different configurations of instances.}
     \Description{None}
   \end{minipage}
\end{figure}

To identify the root cause of prefill queuing, we quantified the system’s prefill processing capacity, defined as the number of prefill tokens computed per second under a given workload.
Figure~\ref{fig/motivation/prefill_capacity} compares the profiled prefill processing capacities of existing methods with a batch size of 16 and a prompt length of 3,000.
We observe that increasing prefill processing capacity leads to a shorter prefill queuing time, from Figures~\ref{fig/motivation/queue} and~\ref{fig/motivation/prefill_capacity}. Specifically, as the PD ratio increases (from 4:4 to 6:2), the prefill processing capacity improves, while the prefill queuing time decreases. However, the maximum prefill processing capacity achieved by PD disaggregation remains lower than that of PD aggregation. This is because, in PD disaggregation, only a subset of instances can contribute to prefill processing capacity, whereas in PD aggregation, all instances are capable of handling prefill tasks.
Additionally, it is worth noting that in PD aggregation, a larger chunk size results in higher prefill processing capacity. This is because computing the same number of prefill tokens requires approximately twice as many iterations for CP512 compared to CP1024, during which roughly twice the number of decode tasks are executed, thereby slowing down the prefill execution speed. This also explains why CP1024 exhibits better TTFT  than CP512 in Figure~\ref{fig/motivation/pd_aggregation_scatter}.

\subsection{Motivations}
To address the dilemma faced by existing methods under balanced TTFT and TPOT SLOs, we propose a \textit{latency-shifting scheduling} paradigm to mitigate the limitations of these two mutually exclusive methods (i.e., PD aggregation and PD disaggregation). The key idea behind it is to shift the latency (i.e., TTFT and TPOT) across prefill/decode phases and across requests.
Specifically, by shifting the latency of requests exceeding SLO constraints to those that significantly satisfy the SLO requirements, we maximize the number of requests that meet the SLO constraints and thereby improve the goodput.
For example, when certain requests show TPOT exceeding constraints (potentially due to the prefill-decode interference), we shift the surplus TPOT to requests with well-satisfied TTFT or TPOT, ensuring all these requests comply with SLO constraints and enhancing overall goodput.

\textbf{Opportunity 1:}
\textit{
The well-satisfied latency of existing methods performs well, leaving substantial room to accommodate shifted latency. 
}

\begin{figure}[t]
    \centering
    \begin{subfigure}[b]{0.23\textwidth}
        \includegraphics[width=\textwidth]{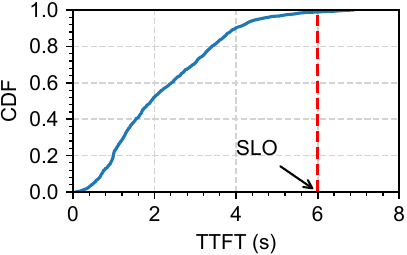}
        \caption{TTFT CDF of PD aggregation (CP1024).}
        \label{fig/key_idea/ttft_cdf_pd_aggregation}
    \end{subfigure}
    \hfill
    \begin{subfigure}[b]{0.23\textwidth}
        \includegraphics[width=\textwidth]{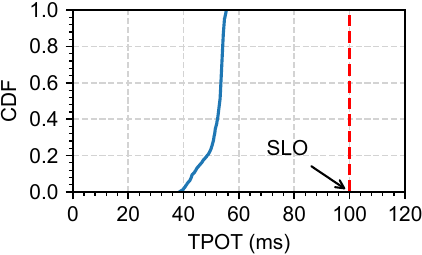}
        \caption{TPOT CDF of PD disaggregation (P6D2).}
        \label{fig/key_idea/tpot_cdf_pd_disaggregation}
    \end{subfigure}
    \caption{Opportunity to shift latency.}
    \label{fig/key_idea/opportunity}
    \Description{None}
\end{figure}

While existing methods excel in only one latency metric (either TTFT or TPOT), their strong performance in their respective domains suggests potential for latency shift.
As shown in Figure~\ref{fig/key_idea/ttft_cdf_pd_aggregation}, over 75\% of requests in PD aggregation achieve a TTFT less than 60\% of the SLO constraint. Similarly, Figure~\ref{fig/key_idea/tpot_cdf_pd_disaggregation} indicates that 100\% of requests in PD disaggregation achieve a TPOT below 60\% of the SLO constraint.
Building on this finding, if requests exceeding SLO constraints can be shifted to these requests with good latency, a significant improvement in goodput can be anticipated.

\textbf{Opportunity 2:}
\textit{Lantecy (TTFT and TPOT) can be shifted across phases by scheduling resources.}

The scheduling policy prioritizes requests for GPU resource (i.e., GPU time) allocation, allowing prioritized requests to be executed first and reducing their latency. If a request prioritized for resources is in the prefill (or decode) stage, it will reduce TTFT (or TPOT).
For example, PD aggregation reduces prefill latency by reserving resources for processing prefill tokens in each iteration using a large chunk size. In contrast, PD disaggregation improves decode latency by allocating all instance resources exclusively to the decode phase.
Notably, if certain requests significantly meet their SLO constraints, it indicates resource over-provisioning for those requests.
Conversely, the latency of requests that are not prioritized for resource allocation will be degraded, thus achieving latency shifting.

Existing methods can implement latency shifting across the prefill and decode phases. For example, increasing the chunk size in PD aggregation demonstrates that TTFT can be shifted to TPOT. This is because increasing the chunk size in PD aggregation will allocate a greater portion of GPU time in each iteration to prefill computation (see Figure~\ref{fig/motivation/iteration_breakdown}). This enhances prefill processing capacity (see Figure~\ref{fig/motivation/prefill_capacity}) and optimizes TTFT (see Figure~\ref{fig/motivation/pd_aggregation_scatter}). 
However, this also reduces the portion of GPU time allocated to decode requests, thereby degrading the TPOT of requests (see Figure~\ref{fig/motivation/pd_aggregation_scatter}).
In contrast, compared to PD aggregation, reserving GPU resources for decode requests, as done in PD disaggregation, allows TPOT to be shifted to TTFT. This prioritization ensures that decode requests obtain resources first, enhancing TPOT at the expense of TTFT.
It is worth noting that PD disaggregation cannot flexibly shift TTFT to TPOT, because the disaggregated decode monopolizes GPU resources, preventing further degradation of TPOT.

In summary, requests prioritized for resource scheduling experience better latency, whereas those scheduled later experience degraded latency. This resource scheduling makes latency shifting possible.

\subsection{Challenges}
\label{subsec/challenges}

However, cross-phase latency shifting alone is insufficient to achieve optimal goodput, as the latency degradation tolerance varies from request to request. The key requirement is to enable latency shifting at the request level, which introduces the following challenges:

\textbf{Challenge 1:} \textit{
Existing methods lack architectural support for request-level latency shifting.
}
Existing methods—whether based on PD aggregation or PD disaggregation—employ uniform GPU instance configurations dedicated either to prefill or decode, resulting in identical service levels for requests across instances. This architectural homogeneity precludes request-level latency optimization or degradation based on their SLO urgency through scheduling, as requests cannot be treated differently.

Specifically, in PD aggregation, all instances share the same configuration. Unifying the chunk size across all instances is both reasonable and efficient.
This is because the TPOT upper bound is determined by requests served in the instance with the largest chunk size, where the prefill-decode interference is highest. If the TPOT of these requests meets the SLO constraints, using smaller chunk sizes in other instances offers no additional benefit. 
However, this uniform configuration prevents architectural support for request-level latency degradation: increasing or decreasing the chunk size affects the TPOT of all requests uniformly. 
Similarly, in PD disaggregation, the prefill instances (affecting TTFT) and decode instances (affecting TPOT) are configured uniformly, respectively. This results in identical service levels of TTFT and TPOT for all requests, leaving no flexibility for scheduling adjustments.

\textbf{Challenge 2:} \textit{ 
Request-level TPOT degradation is hindered by batch processing and output length uncertainty.
}
First, batch processing makes request-level TPOT degradation seemingly impossible, as the degradation must occur at the batch level. 
Batch processing is a crucial optimization that improves throughput and resource utilization by allowing multiple requests to share the overhead of model loading. However, this shared processing creates interdependence, where a performance change intended for one request inevitably affects all co-located requests, regardless of whether they benefit from or can tolerate such changes. 
For example, using chunked prefill with a larger chunk size can shift the TTFT of a prefill request to the TPOT of a decode request that can tolerate increased TPOT. Nevertheless, this approach may also cause unintended TPOT degradation for other decode requests within the batch that could not tolerate such degradation.
This lack of isolation fundamentally limits the system’s ability to selectively degrade TPOT at the request level.

\begin{figure}
    \centering
    \includegraphics[width=0.3\textwidth]{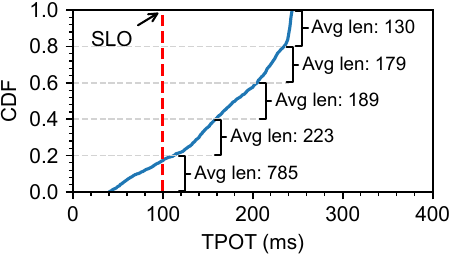}
    \caption{Relationship between TPOT and decode length for CP1024.}
    \label{fig/challenge_tpot_cdf}
    \vspace{-10pt}
    \Description{None}
\end{figure}

Second, the unpredictability of output lengths prevents the system from identifying which decode requests are safe and suitable to degrade. 
Short-output requests are more vulnerable to PD interference, leading to excessive TPOT degradation, because the output length acts as the denominator in calculating interference intensity (as defined in \S~\ref{subsec/motivation/PD_aggregation}). This effect is evident in our experiments, as shown in Figure~\ref{fig/challenge_tpot_cdf}.
Consequently, it is necessary to control TPOT degradation for requests with short output lengths. 
However, the output lengths cannot be predetermined until the \textit{end-of-sequence} token is generated in auto-regressive models. This uncertainty impedes the determination of which requests to degrade and by how much, making it challenging for the scheduler to proactively perform request-level TPOT degradation throughout the decode process.
Although some~\cite{OutputLengthPrediction1, OutputLengthPrediction2,OutputLengthPrediction3} works have attempted to predict output lengths, achieving high prediction accuracy across all datasets remains very difficult; for example, these works can only reach an accuracy of 60\%–81\% on certain datasets.
An x\% prediction error rate can result in an equivalent x\% decrease in SLO attainment, thereby reducing the cost efficiency of the LLM serving system. Therefore, output length prediction with insufficient accuracy is not suitable for deployment in a production LLM serving environment.

\textbf{Challenge 3:}
\textit{
Request-level TTFT degradation is non-trivial, as it requires jointly considering both the request's execution and queuing time.}
Long prefill requests inherently have longer execution times and are more likely to violate TTFT constraints, making them unsuitable for further degradation.
Recent studies~\cite{WorkloadStudy, WildChat} show a broad distribution of prefill lengths: from very short to very long. 
For example, assuming prefill lengths mostly range from 2k to 16k tokens (as in the Arxiv summarization dataset~\cite{Arxiv}) and a prefill throughput of 5k tokens per second (consistent with Figure 8), the prefill execution time varies from 0.4s to 3.2s. If the TTFT SLO is set to 3.5s, longer requests already approach this limit, necessitating the avoidance of TTFT degradation. Conversely, short requests are more suitable for accepting shifted TTFT because they can tolerate slower execution.
Analogously, requests that have already endured long queuing delays have consumed a significant portion of their TTFT budget. This leaves them with little tolerance for additional execution latency, making it crucial to protect them from further performance degradation.
Overall, implementing TTFT degradation for individual requests is not straightforward.

\section{The Design of \sysname}
\label{sec/design}

In this section, we first present the architectural overview of \sysname, which unifies PD aggregation and disaggregation within a single flexible framework (\S~\ref{subsec/design/overview}). We then introduce \textit{hybrid-mode inference} (\S~\ref{subsec/design/hybrid}), a capability unique to \sysname that enables latency shifting to achieve superior goodput. Finally, we describe two targeted scheduling mechanisms that support hybrid-mode inference: \textit{flowing decode scheduling} for controlling TPOT (\S~\ref{subsec/design/decode_scheduling}) and \textit{length-aware prefill scheduling} for managing TTFT (\S~\ref{subsec/design/prefill_scheduling}).

\subsection{Architectural Overview}
\label{subsec/design/overview}

We present \sysname, an LLM serving system that unifies PD disaggregation and aggregation to achieve goodput-optimal performance under any combination of TTFT and TPOT SLOs. \sysname is built on a unified aggregation-disaggregation architecture composed of differentiated-capability instances, i.e., P-heavy and D-heavy instances. On top of this architecture, \sysname exposes three configurable sliders to control the ratio between prefill-heavy and decode-heavy instances, and the chunk sizes for each. By adjusting these sliders, \sysname can dynamically adapt to a wide range of SLO regimes. 

\begin{figure}[t]
    \centering
    \includegraphics[width=\columnwidth]{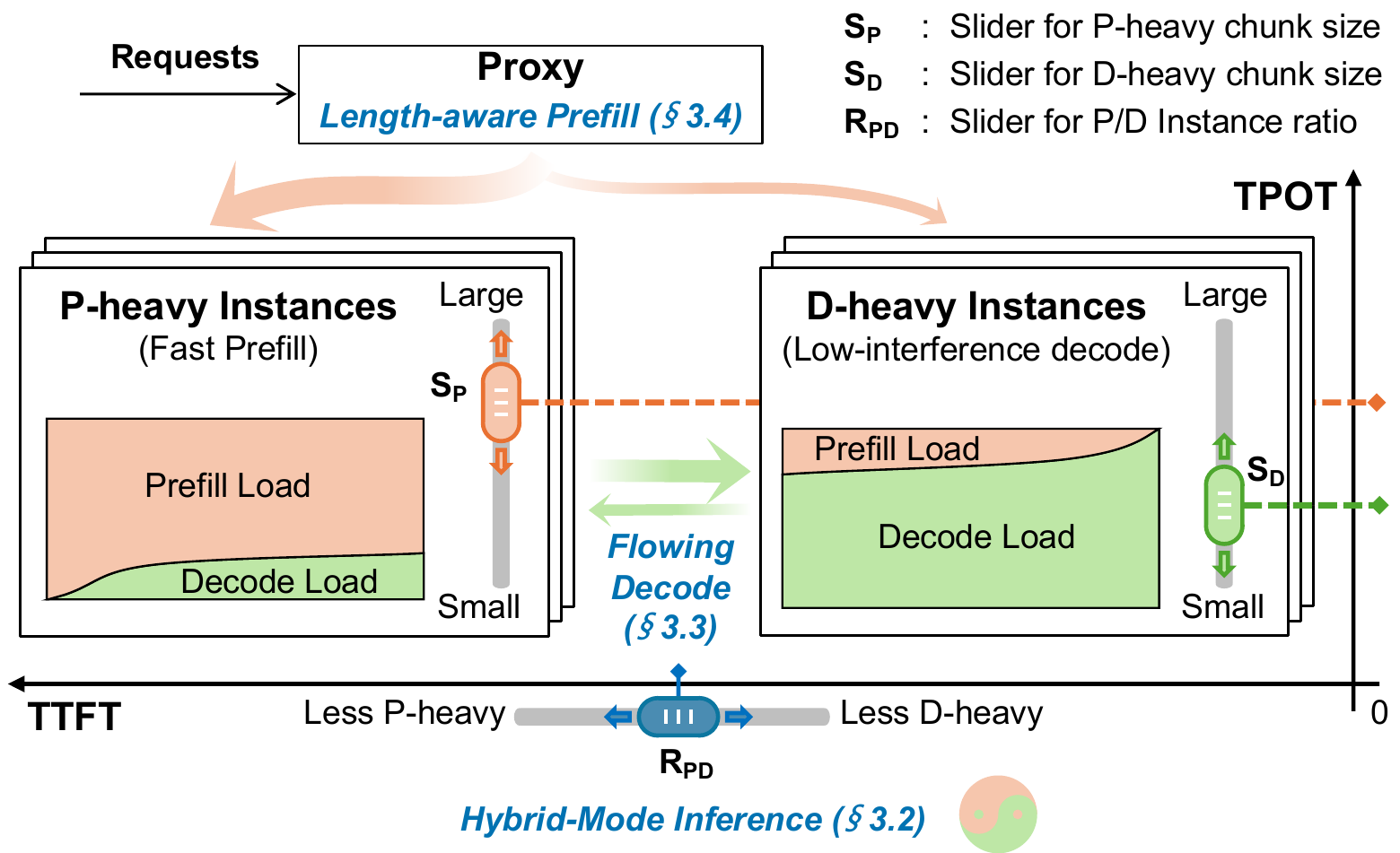}
    \caption{The system overview of \sysname. The system configures instances as either prefill-heavy or decode-heavy using different chunk sizes, each offering differentiated capability. A length-aware proxy routes requests to an appropriate instance for their prefill phase (either fast or degraded). The subsequent decode execution then "flows" between the two instance types to dynamically manage the TPOT (optimized or degraded).}
    \label{fig/design_overview}
\end{figure}

\textbf{Differentiated-Capability Instances}. As illustrated in Figure~\ref{fig/design_overview}, \sysname comprises a proxy and multiple inference instances. The proxy orchestrates the execution of prefill and decode tasks by dispatching them to appropriate instances based on their capabilities. The inference instances in \sysname are divided into two types:

\begin{itemize}
    \item \textit{P-heavy} instances are optimized for prefill tasks. They are configured with \textit{larger chunk sizes}, enabling them to efficiently process compute-intensive prefill workloads and minimize TTFT. However, when handling decode tasks, they suffer from \textit{high prefill-decode interference}, leading to higher TPOT.

    \item \textit{D-heavy} instances are optimized for decode tasks. Configured with \textit{smaller chunk sizes}, they provide low-interference decode, thereby reducing TPOT. Although slower at prefill, D-heavy instances can still handle certain degradable prefill tasks, improving utilization over pure PD disaggregation.
\end{itemize}

This capability differentiation—achieved purely through chunk size configuration—allows \sysname to flexibly combine the strengths of PD aggregation and disaggregation within a single unified framework.

\textbf{Configurable Sliders.} \sysname introduces three sliders to navigate the PD design space: 
\begin{itemize}
\item ${R}_{PD}$: the ratio of P-heavy to D-heavy instances.
\item $S_P$: the chunk size for executing chunked prefill in P-heavy instances.
\item $S_D$: the chunk size for executing chunked prefill in D-heavy instances.
\end{itemize}
Larger chunk sizes improve an instance’s prefill throughput, which benefits TTFT, but also increase prefill-decode interference, thereby degrading TPOT (\S~\ref{subsec/motivation/PD_aggregation}). Similarly, increasing ${R}_{PD}$ enhances the system's overall prefill processing capacity by allocating more P-heavy instances, but reduces the availability of decode resources (\S~\ref{subsec/motivation/PD_disaggregation}).

By adjusting these sliders, \sysname tunes the system's latency profile to meet different combinations of TTFT and TPOT SLOs. The optimal configuration for a given workload and SLO can be determined via offline search, following approaches from prior work~\cite{ChunkedPrefill,DistServe,SplitWise}. For example, under a strictly tight TPOT constraint, \sysname can be configured as a pure PD disaggregation system by setting $S_D$ to exclude the prefill tokens and assigning $S_P$ to the maximum content length, effectively disabling chunked prefill. An appropriate ${R}_{PD}$ is then selected based on the workload characteristics. Conversely, when TTFT is the primary constraint, \sysname can operate as a pure PD aggregation system by setting $S_D = S_P$ to a common chunk size across all instances.
Under balanced TTFT and TPOT constraints, \sysname provides a \textit{hybrid mode} that enables latency shifting to achieve optimal goodput, as described in the next subsection.
To adapt to dynamic workloads, our system adopts an on-demand search-and-reconfigure strategy like DistServe, which triggers a new search and reconfiguration only upon significant workload changes. This reconfiguration completes in minutes and is far shorter than the typical hourly-scale shifts in workload patterns~\cite{DistServe}.

\subsection{Hybrid-Mode Inference}
\label{subsec/design/hybrid}

To support balanced TTFT and TPOT SLOs and maximize goodput, \sysname introduces \textit{hybrid-mode inference}, which enables \textit{latency shifting}—strategically redistributing latency (i.e., TTFT and TPOT) both across the prefill and decode phases and across requests—by leveraging the differentiated capabilities of instances in \sysname (\S~\ref{subsec/design/overview}). 

Unlike traditional PD aggregation or disaggregation, hybrid-mode inference uniquely combines two complementary scheduling principles: \textit{aggregated batch handling} for high resource efficiency and \textit{disaggregated request handling} for fine-grained latency control, as summarized in Table~\ref{tab/policy-comparison}. This design enables \sysname to balance latency and throughput under diverse SLO regimes.

\textbf{Aggregated Batch Handling for High Utilization}. Inspired by PD aggregation, this hybrid mode allows both P-heavy and D-heavy instances to process mixed batches containing both prefill and decode tasks. By enabling all instances to contribute to prefill processing and piggyback decode requests with chunked prefill, this approach boosts the system's total prefill throughput and improves the overall GPU utilization.

\textbf{Disaggregated Request Handling for Fine-Grained Control}. In line with PD disaggregation, this hybrid mode allows the prefill and decode phases of a single request to be executed on different instances. This enables fine-grained, per-request latency optimization or degradation. 
For example, a request's prefill phase can be routed to a P-heavy instance to ensure low TTFT, while its decode phase can be assigned to a D-heavy instance for low TPOT. Conversely, a request can be deliberately degraded by assigning its prefill to a D-heavy instance and its decode to a P-heavy instance, freeing specialized resources for latency-critical requests. This mechanism forms the core of \sysname’s latency shifting capability.

In effect, hybrid-mode inference unifies the scheduling flexibility of PD disaggregation with the high utilization efficiency of PD aggregation. More importantly, it provides the necessary foundation for the latency-shifting scheduling strategies described in \S~\ref{subsec/design/decode_scheduling} and \S~\ref{subsec/design/prefill_scheduling}.

\subsection{Flowing Decode Scheduling}
\label{subsec/design/decode_scheduling}

\begin{figure}[t]
    \centering
    \includegraphics[width=\columnwidth]{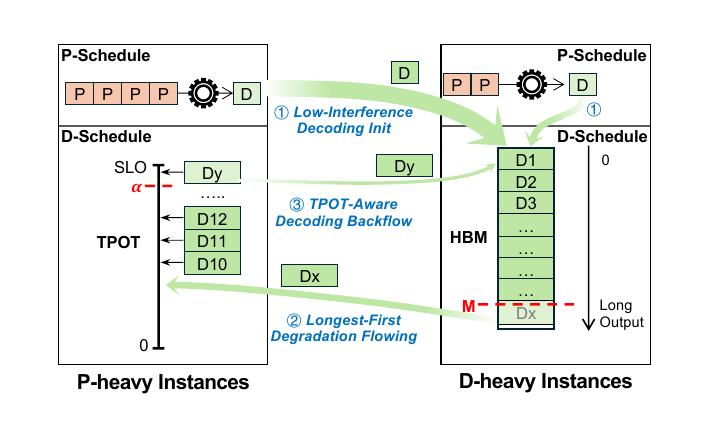}
    \caption{An illustration of flowing decode scheduling. The scheduler manages TPOT by \ding{172} starting decode requests on D-heavy instances, \ding{173} offloading the longest ones to P-heavy for TPOT degradation, and \ding{174} returning any that approach their TPOT SLO.}
    \label{fig/design/flowing_decode_scheduling}
    \Description{None}
\end{figure}

To enable request-level TPOT degradation, we introduce \textit{flowing decode scheduling} for fine-grained, per-request latency control. The core mechanism involves dynamically migrating decode requests between D-heavy instances and P-heavy instances. 
This migration allows the system to intentionally and selectively degrade the TPOT of certain requests, thereby reallocating over-provisioned resources to other requests that require lower latency, without the constraints of batch processing or the need for pre-determined output lengths. 
In contrast to existing methods, our approach avoids the consistent high interference characteristic of PD aggregation while enabling the dynamic TPOT degradation that PD disaggregation lacks.
As illustrated in Figure~\ref{fig/design/flowing_decode_scheduling}, the flowing decode process encompasses the following three key stages.

\textbf{\ding{172} Low-Interference Decode Init:}
After a request completes its prefill phase, it is initially scheduled to a D-heavy instance to begin low-interference decode and prevent premature TPOT violations. This strategy is necessary because if a short-output request (e.g., producing only 2 output tokens, but this is unknown a priori) begins decode on a P-heavy instance with high interference, it may complete decode there and violate its TPOT constraint.
The selection of the initial D-heavy instances considers both load balancing and the minimization of KV cache transfers. If the prefill of a request is executed on a P-heavy instance, the proxy schedules it to the D-heavy instance with the lowest decode load (i.e., HBM usage). Conversely, if the prefill stage of a request is executed on a D-heavy instance, it will perform the in-place decode to minimize KV cache transfers between instances.

\textbf{\ding{173} Longest-First Degradation Flowing:}
When the HBM capacity of D-heavy instances reaches a predefined memory watermark $M$ (e.g., 95\% utilization), we selectively offload a portion of decode requests to P-heavy instances, thereby degrading their TPOT to release GPU resources for other requests.
The selection of requests for degradation is strategic: to avoid penalizing interference-sensitive short-output jobs, whose lengths are unknown a priori (Challenge 2), we innovatively prioritize offloading requests with the current longest output. 
These requests are ideal candidates for offloading, as they have already benefited from numerous iterations on low-interference D-heavy instances and can thus better absorb the performance degradation. 
Notably, the predefined memory watermark $M$ is essential to guarantee sufficient memory is reserved to accept at least one new decode request.

This approach deliberately degrades the most degradable TPOT requests to: (1) free resources on D-heavy instances, optimizing TPOT for new decode requests; and (2) limit resource usage on large-chunk P-heavy instances, improving TTFT for prefill requests.
The degrading flowing method effectively addresses the technical challenge of request-level TPOT degradation in batch processing. This is because it isolates requests requiring degradation from the original batch (on D-heavy instances) and reassigns them to high-interference batches (on P-heavy instances), thereby preventing interference diffusion to the other requests in the original batch.

\textbf{\ding{174} TPOT-Aware Decode Backflow:}
To prevent excessive interference for decode requests in P-heavy instances, we monitor their real-time TPOT and migrate those approaching the TPOT SLO to D-heavy instances.
Specifically, a request is considered to be approaching its SLO when its current TPOT surpasses the product of the SLO value and a predefined \textit{approaching factor} $\alpha$ (e.g., 0.96 in our experiments), enabling proactive optimization before SLO violations occur.
Upon flowing back, the decode request is logically treated as a new request, with its output length reset. This reset accounts for the neutralization between the low interference on D-heavy instances and the high interference on P-heavy instances, ensuring the current TPOT closely adheres to the SLO constraint. 
Each backflow typically triggers a degrading flowing in D-heavy instances to release HBM capacity. Consequently, this shifts the TPOT from the request in the backflow to other requests in the degrading flowing.

However, we intend for optimizing flowing to serve as a safeguard mechanism for TPOT degradation rather than a frequent occurrence. This is because the triggered degrading flowing may prematurely degrade the TPOT of certain requests, reducing their degradation margin. In fact, the frequent backflow indicates improper architectural configuration, where excessive TTFT optimization compromises TPOT, overloading the decode scheduling. For example, the underlying causes may include: (1) oversized chunk size settings in either D-heavy or P-heavy instances, or (2) insufficient quantity of D-heavy instances. Thus, to avoid the frequent backflow, the system architecture should be adjusted to better match workload characteristics.

\begin{algorithm}[t!]
\caption{Decode Scheduling Algorithm in Instances}
\label{alg/decode}
\begin{algorithmic}[1]
\Statex \textbf{Input:} Decode request set $S$, TPOT SLO $\tau_{tpot}$, Current memory usage $m$, Instance type $type$, Approach factor $\alpha$, Memory watermark $M$
\Statex \textbf{Output:} Optimizing set $O$ or Degrading set $D$
\If{$type$ is \texttt{P-heavy}}
    \State $O \gets \{ r \in S \mid r_{\text{tpot}} > \tau_{tpot} * \alpha \}$ \Comment{Approaching SLO}
    \State \Return $O$
\ElsIf{$type$ is \texttt{D-heavy}}
    \State $D \gets \emptyset$
    \State $m_\text{release} \gets 0$ \Comment{Memory size to release}
    \While{$m - m_{\text{release}} > M$}
        \State Select $r^* \gets \arg\max_{r \in S \setminus D} (r_{\text{current\_output\_len}})$
        \State $D \gets D \cup \{ r^* \}$
        \State $m_{\text{release}} \gets m_{\text{release}} + r^*_{\text{memory}}$
    \EndWhile
    \State \Return $D$
\EndIf
\end{algorithmic}
\end{algorithm}

Algorithm~\ref{alg/decode} details the request selection mechanism for backflow and degrading flowing. Implemented in the instance scheduler, this algorithm is invoked during the scheduling phase of each inference iteration.
It evaluates the current states of instances and requests, together with the TPOT SLO, to select backflow requests for P-heavy instances and degrading requests for D-heavy instances.
In Lines 1-3, the scheduler in P-heavy instances selects requests to conduct backflow. Line 2 calculates each request's current TPOT value and adds those nearing the SLO to the optimizing set. 
Lines 4–12 handle D-heavy instances: if memory usage exceeds the threshold M, the scheduler repeatedly selects the longest decode request currently (Line 8), adds it to the degrading set (Line 9), and updates the released memory (Line 10) until usage drops below M (Line 7).
Finally, the algorithm outputs the chosen optimizing or degrading set. The decode requests within the set are subsequently distributed to the D-heavy or P-heavy instances to optimize or degrade TPOT through the proxy in a load-balanced manner.

\subsection{Length-aware Prefill Scheduling}
\label{subsec/design/prefill_scheduling}

To enable request-level TTFT degradation, we propose a length-aware prefill scheduling strategy. 
The key idea is to exploit the lower urgency of short prefill requests by routing them to D-heavy instances, intentionally slowing their execution to free up P-heavy instances for more time-sensitive, long prefill requests.

\begin{figure}[t!]
    \centering
    \includegraphics[width=0.48\textwidth]{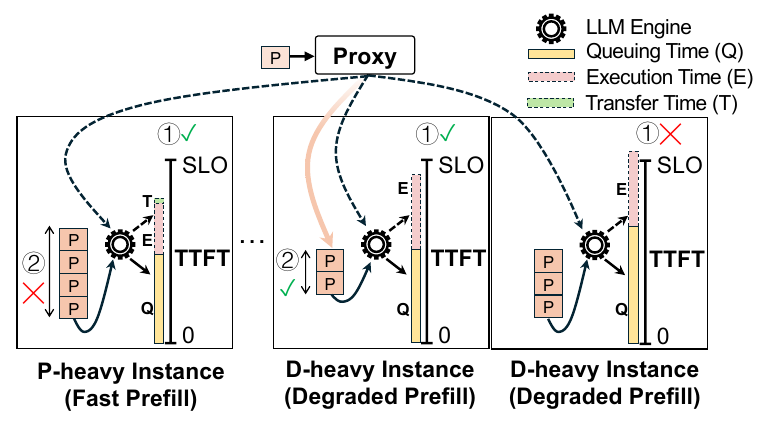}
    \caption{An illustration of length-aware prefill scheduling. The scheduler preferentially assigns short prefill requests to slower D-heavy instances when their estimated TTFT (queuing + execution + transfer time) satisfies the SLO. This reserves the faster P-heavy instances for long, more time-sensitive requests to ensure they also meet their SLOs.}
    \label{fig/design/length_aware_prefill_schduling}
    \Description{None}
\end{figure}

Figure~\ref{fig/design/length_aware_prefill_schduling} illustrates the scheduling strategy of the prefill algorithm. This algorithm operates within the proxy, assigning each newly arrived prefill request to an instance. 
The scheduling involves two steps: first, identifying \textit{feasible instances} where assigning the request will not violate its TTFT SLO constraint; second, selecting among feasible instances with the fewest queued prefill tokens, typically favoring a D-heavy instance. The proxy then enqueues the request in the prefill queue of the selected instance, where requests are processed in a first-come, first-served manner.

Specifically, in the first step, the proxy estimates the TTFT for the incoming prefill request on each instance. 
For D-heavy instances, TTFT is the sum of queuing time (Q) and execution time (E); for P-heavy instances, transfer time (T) is also included due to the need to transfer the KV cache.
The queuing time for a request on an instance is defined as the total estimated execution time of the remaining prefill tasks on the instance. Accurately estimating the execution time of a prefill request requires modeling factors such as request length, instance configuration, and batch information. The recent research Vidur~\cite{Vidur} models it and provides an accurate and efficient execution time predictor, which we leverage to estimate both queuing time (Q) and execution time (E). According to our experiments, this predictor completes estimation within negligible tens of microseconds.
The transfer time (T) is determined by the KV cache size to transfer and link bandwidth, but is typically negligible under high-speed interconnects, as discussed in \S ~\ref{subsec/background/scheduling_policies}.
After estimating TTFTs, instances that can process the prefill request within the TTFT SLO are selected as feasible instances. 

In the second step, the scheduler selects the instance with the fewest queuing prefill tokens. This approach is motivated by two reasons. 
First, if a D-heavy instance is among the feasible instances, it is highly likely to be selected, ensuring that degradable short requests are preferentially degraded. This is because D-heavy instances, having lower prefill processing capacity than P-heavy instances, accommodate fewer queuing tokens under the same TTFT constraint. 
Second, this strategy helps balance load: if a P-heavy instance has fewer tokens than all feasible D-heavy instances (e.g., after multiple degradable prefill tasks have been assigned to D-heavy instances), the request is assigned to the P-heavy instance, thereby avoiding load imbalance. In this scenario, degradation is unnecessary since a less-loaded P-heavy instance is available for fast prefill.

It is worth noting that if the feasible instance set is empty, the request will inevitably violate the TTFT SLO, often due to a sudden surge in prefill workload. Prior work has proposed the \textit{early rejection} strategy~\cite{Mooncake} to proactively drop such requests, thus preventing instance overload and subsequent cascading SLO violations. However, to ensure a fair comparison under identical load conditions with PD aggregation—which generally provides sufficient prefill processing capacity—we randomly assign such requests to an instance in our experiments, even though this may occasionally violate the TTFT SLO.

\begin{algorithm}[t]
\caption{Prefill Scheduling Algorithm in the Proxy}
\label{alg/prefill}
\begin{algorithmic}[1]
\Require
    Request $r$,
    Instance set $\mathcal{I}$,
    TTFT SLO $\tau_{ttft}$
\Ensure
    Scheduled instance $i^*$ or $\emptyset$
    
\State $\mathcal{I}' \gets \emptyset$
\For{each $i \in \mathcal{I}$}

    \State $Q \gets \sum_{r' \in i.\text{queue}} \text{Estimate}\left(r'.\text{len},\ i.\text{chunk},\ i.\text{batch}\right)$

    \State $E \gets \text{Estimate}(r.\text{len}, i.\text{chunk}, i.\text{batch})$
    \State $T \gets \mathbb{I}\{i_{type} = \text{P-heavy}\} \cdot \frac{r_{\text{transfer\_size}}}{link\_bw}$

    \If{$Q + E + T < \tau_{ttft}$}
        \State $\mathcal{I}' \gets \mathcal{I}' \cup \{ i \}$
    \EndIf
\EndFor

\If{$\mathcal{I}' \neq \emptyset$}
    \State $i^* \gets \argmin_{i \in \mathcal{I}'} \sum_{r' \in i.\text{queue}} r'.\text{len}$
    \State \Return $i^*$
\Else
    \State \Return $\emptyset$
\EndIf

\end{algorithmic}
\end{algorithm}

Algorithm~\ref{alg/prefill} presents the detailed implementation of the length-aware prefill scheduling. The algorithm takes three input parameters: a newly arrived request, information about all instances, and the TTFT SLO constraint. It outputs a suitable instance capable of processing the prefill request within the TTFT constraint, if such an instance exists.
Lines 1-9 identify the set of feasible instances capable of processing the request within the TTFT SLO. An instance is considered feasible if the sum of its queuing time (Line 3), the execution time for the new request (Line 4), and the potential transfer time does not exceed the TTFT SLO (Line 5). 
Specifically, both queuing time and execution time are estimated using an execution time model (e.g., the execution time predictor in Vidur).
For only P-heavy instances, the transfer time is determined by dividing the required transfer size by the link bandwidth.
Lines 10–12 select the final feasible instance with the fewest queuing prefill tokens to handle the request, considering both TTFT degradation and load balancing.
Typically, it is a D-heavy instance, which strategically allows for TTFT degradation. 
Lines 13-15 indicate that if no instance can execute the request within the TTFT constraint, the algorithm returns an empty result.

\subsection{Implementation}
\label{subsec/implementation}

We implement \sysname on the open-source vLLM~\cite{vllm_project}.
We employ the chunked prefill implementation from vLLM for our P-heavy and D-heavy instances with different chunk sizes.
Regarding the KV transfer between the P-heavy instances and the D-heavy instances, we extend vLLM's KV transfer module to enable mutual communication between any two instances via NCCL~\cite{NCCL}. To enhance the efficiency of the KV transfer, we decouple the transfer from the critical path of the model execution in vLLM, making it asynchronous. Additionally, we utilize fused CUDA operators to store the received KV cache into vLLM's paged memory, thereby reducing its CPU overhead.
\zh{

}

\section{Performance Evaluation}
\label{sec/evaluation}

\subsection{Experiment Setup}

\textbf{Cluster Testbed and Models} 
We deploy our experiments on 8 NVIDIA SXM A100-80GB GPUs connected via NVLINK, using the widely adopted Qwen2.5 series models with FP16 precision. Due to single-node limitations, our experiments focus on multiple instances of Qwen2.5-14B and Qwen2.5-32B. 
To keep the 32B model’s HBM usage below half of total capacity and ensure sufficient KV cache space, we apply tensor parallelism (TP) and set TP=2 for Qwen2.5-32B.

\begin{figure}
    \centering
    \begin{subfigure}[b]{0.23\textwidth}
        \includegraphics[width=\textwidth]{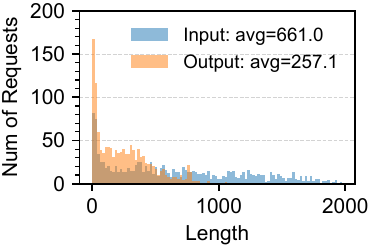}
        \caption{ShareGPT}
        \label{fig/sharegpt_hist}
    \end{subfigure}
    \hfill 
    \begin{subfigure}[b]{0.23\textwidth}
        \includegraphics[width=\textwidth]{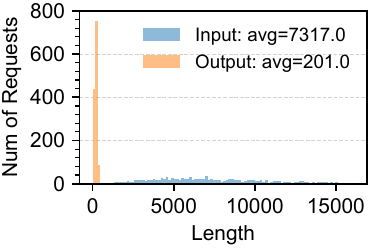}
        \caption{Arxiv Summarization}
        \label{fig/arxiv_hist}
    \end{subfigure}
    \caption{The input and output length distributions of ShareGPT and Arxiv Summarization datasets.}
    \label{fig/length_distribution}
    \Description{None}
\end{figure}

\begin{table}
\caption{Evaluated workloads and SLO constraints. }
\label{tab/slo_constraints}
\centering
\begin{tabular}{lcc}
\toprule
\textbf{(TTFT, TPOT)} & \textbf{SLO1} & \textbf{SLO2} \\
\midrule
\textbf{ShareGPT} & (3s, 110ms) & (4s, 70ms) \\
\textbf{Arxiv Summarization} & (4s, 70ms) & (6s, 50ms) \\
\bottomrule
\end{tabular}
\end{table}

\textbf{Workloads Setup.}
To simulate real-world serving scenarios, we assess a chatbot~\cite{ChatGPT} and a summarization application~\cite{Summarization}, following the methodology of~\cite{DistServe,ChunkedPrefill}. The chatbot uses the \textit{ShareGPT} dataset~\cite{ShareGPT}, comprising user-shared ChatGPT conversations, while summarization experiments use the \textit{ArXiv Summarization} dataset~\cite{Arxiv}, characterized by long prefill sequences. Since the datasets lack timestamp information, we simulate request arrivals using a Poisson process with varying rates, as in prior work~\cite{DistServe,ChunkedPrefill}. Figure~\ref{fig/length_distribution} presents the input and output length distributions. We filter outliers by discarding ShareGPT requests exceeding 2048 tokens and ArXiv Summarization requests over 16,384 tokens.

We evaluate performance under two balanced SLO configurations to highlight our design's benefits across varied user-defined requirements: SLO1 (relatively lower TTFT and higher TPOT) and SLO2 (relatively higher TTFT and lower TPOT), as detailed in Table~\ref{tab/slo_constraints}. Summarization tasks typically use longer prefill prompts and demand faster output than chatbot tasks, resulting in higher TTFT but lower TPOT constraints, in line with previous studies~\cite{DistServe}. For Qwen2.5-32B, all TPOT SLOs are relaxed by 10 ms to accommodate increased execution time and communication overhead from tensor parallelism.

\textbf{Metrics.}
Following prior work~\cite{DistServe}, we evaluate all methods based on the maximum achievable goodput under the 90\% SLO attainment rate (\S \ref{subsec/evaluation/end-to-end}). Additionally, we demonstrate the reduction in bottleneck latency compared to existing approaches at the maximum achievable goodput of \sysname, highlighting the direct cause of the improved goodput (\S \ref{subsec/evaluation/latency_reduction}. Moreover, we conduct the performance breakdown to observe changes in both latency and SLO attainment rate, thereby demonstrating the effectiveness of our proposed approach (\S \ref{subsec/evaluation/breakdown}). Finally, we present the analysis of the overhead introduced by our design (\S \ref{subsec/evaluation/overhead_analysis}).

\begin{figure*}[t!]
    \centering
    \hspace*{2em}\includegraphics[width=0.5\linewidth]{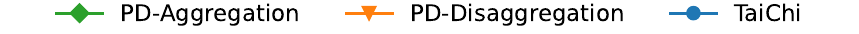} \\ 
    \begin{subfigure}[b]{0.24\textwidth}
        \includegraphics[width=\textwidth]{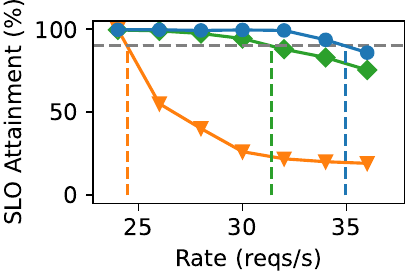}
        \caption{14B under SLO1} 
        \label{fig/evaluation/end-to-end/chatbot_goodput/sub1}
    \end{subfigure}
    \hfill 
    \begin{subfigure}[b]{0.24\textwidth}
        \includegraphics[width=\textwidth]{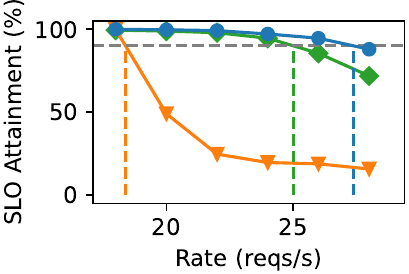}
        \caption{32B under SLO1} 
        \label{fig/evaluation/end-to-end/chatbot_goodput/sub2}
    \end{subfigure}
    \hfill 
    \begin{subfigure}[b]{0.24\textwidth}
        \includegraphics[width=\textwidth]{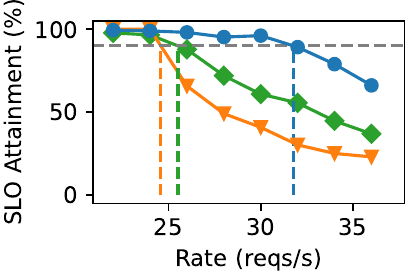}
        \caption{14B under SLO2} 
        \label{fig/evaluation/end-to-end/chatbot_goodput/sub3}
    \end{subfigure}
    \hfill 
    \begin{subfigure}[b]{0.24\textwidth}
        \includegraphics[width=\textwidth]{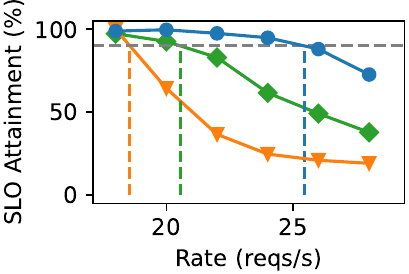}
        \caption{32B under SLO2} 
        \label{fig/evaluation/end-to-end/chatbot_goodput/sub4}
    \end{subfigure}
    \hfill 
    \caption{Goodput (vertical lines) for chatbot tasks using Qwen-2.5 models under SLO1 and SLO2.}
    \label{fig/evaluation/end-to-end/chatbot_goodput}
    \Description{None}
\end{figure*}

\begin{figure*}[t!]
    \centering
    \begin{subfigure}[b]{0.24\textwidth}
        \includegraphics[width=\textwidth]{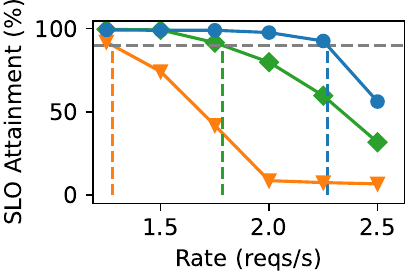}
        \caption{14B under SLO1} 
        \label{fig/evaluation/end-to-end/summarization_goodput/sub1}
    \end{subfigure}
    \hfill 
    \begin{subfigure}[b]{0.24\textwidth}
        \includegraphics[width=\textwidth]{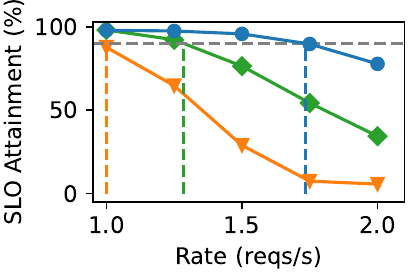}
        \caption{32B under SLO1} 
        \label{fig/evaluation/end-to-end/summarization_goodput/sub2}
    \end{subfigure}
    \hfill 
    \begin{subfigure}[b]{0.24\textwidth}
        \includegraphics[width=\textwidth]{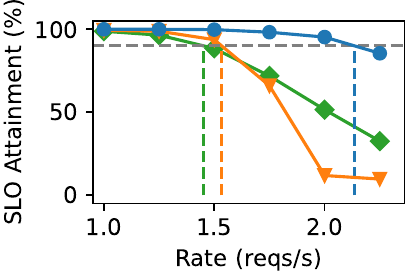}
        \caption{14B under SLO2} 
        \label{fig/evaluation/end-to-end/summarization_goodput/sub3}
    \end{subfigure}
    \hfill 
    \begin{subfigure}[b]{0.24\textwidth}
        \includegraphics[width=\textwidth]{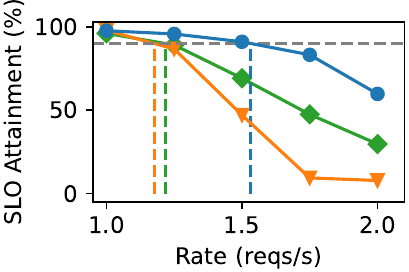}
        \caption{32B under SLO2} 
        \label{fig/evaluation/end-to-end/summarization_goodput/sub4}
    \end{subfigure}
    \hfill 
    \caption{Goodput (vertical lines) for summarization tasks using Qwen-2.5 models under SLO1 and SLO2.}
    \label{fig/evaluation/end-to-end/summarization_goodput}
    \Description{None}
\end{figure*}

\textbf{Baseline.} We compare \sysname to baseline systems, which have both been implemented in the vLLM project~\cite{vllm_project}:

\begin{itemize}
    \item \textbf{PD aggregation.} 
Chunked prefill, as a representative of PD aggregation, divides the prefill tasks into small chunks to improve hardware resource utilization and ensure that decode tasks are not stalled for too long. It mitigates but cannot completely eliminate prefill-decode interference caused by long prompts. We prioritize setting its chunk size to meet the required prefill processing capacity for tested workloads, preventing the prefill requests queue from growing excessively and causing TTFT explosion. 
We also show the latency performance of prioritizing the bounded TPOT with a small-chunk configuration in Section~\ref{subsec/evaluation/breakdown} for a comparison.

    \item \textbf{PD Disaggregation.}
To optimize TPOT, PD disaggregation reserves dedicated instances for decode to eliminate prefill-decode interference issues. Since decode occupies some instances, PD disaggregation faces insufficient prefill processing capacity. We set the PD ratio to the configuration that yields the best TTFT performance, as its TPOT consistently performs well.
We extend the PD disaggregation functionality in the vLLM project~\cite{vllm_project}, transforming its original one-to-one KV cache transmission into a many-to-many KV Cache transmission via NCCL~\cite{NCCL}.    

\end{itemize}

\subsection{End-to-end Experiments}
\label{subsec/evaluation/end-to-end}

In this section, we compare \sysname\ with the baseline on real-world application datasets. \sysname\ increases maximum goodput by 9–47\% over PD aggregation and 29–77\% over PD disaggregation across diverse workloads, while maintaining 90\% SLO compliance.

\textbf{Chatbot.} 
We evaluate the performance of \sysname on the chatbot application using Qwen2.5 models as Figure~\ref{fig/evaluation/end-to-end/chatbot_goodput} shows.
Increasing the request rate leads to higher latency violations, reducing SLO attainment. The vertical line indicates the maximum request rate that maintains latency compliance for over 90\% of requests. 

Compared to PD aggregation, \sysname achieves 9–11\% and 24–25\% higher goodput under SLO1 and SLO2, respectively. This improvement is due to \sysname’s ability to maintain similar prefill processing capacity while effectively bounding TPOT through flowing decode scheduling, which assigns all decode requests first to low-interference D-heavy instances.
For SLO1, we use two P-heavy instances (chunk size 1024) and two D-heavy instances (chunk size 512). In contrast, the chunked prefill approach requires all four instances to use a chunk size of 1024 to match prefill processing capacity and TTFT, but this increases interference and causes TPOT violations.
For SLO2, we configure two P-heavy instances (chunk size 1024) and two D-heavy instances (chunk size 128), reducing D-heavy chunk size to tighten TPOT and lower prefill processing capacity, thereby taking advantage of the relaxed TTFT SLO. PD aggregation, by contrast, needs four instances with chunk size 512 for similar prefill processing capacity, but this significantly violates the TPOT constraint.

Compared to PD disaggregation, our design increases goodput by 43–49\% under SLO1 and 29–37\% under SLO2. This improvement is due to our D-heavy instances, which supplement prefill processing capacity that PD disaggregation lacks. In PD disaggregation, two instances each are dedicated to prefill and decode, as decode requires the HBM of two instances. In contrast, our approach allows D-heavy instances to assist with prefill, enhancing overall prefill processing capacity and supporting higher QPS without violating TTFT constraints. Additionally, higher QPS leads to more concurrent decode requests; our solution routes these to P-heavy instances without needing extra decode instances, unlike PD disaggregation.

\begin{figure}[t!]
    \centering
    \begin{subfigure}[b]{0.40\textwidth}
        \includegraphics[width=\textwidth]{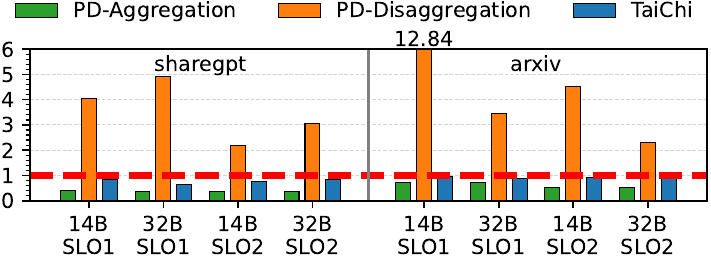}
        \caption{TTFT normalized to the SLO}
        \label{fig/evaluation/ttft_reduction}
    \end{subfigure}
    \begin{subfigure}[b]{0.40\textwidth}
        \includegraphics[width=\textwidth]{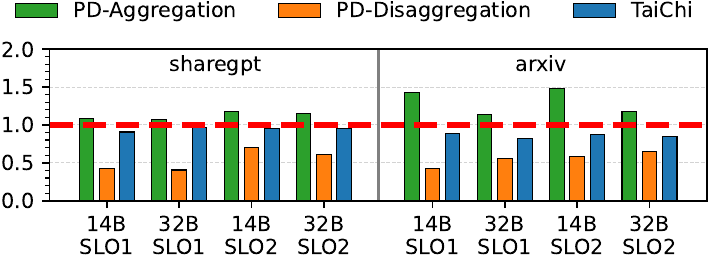}
        \caption{TPOT normalized to the SLO}
        \label{fig/evaluation/tpot_reduction}
    \end{subfigure}
    \caption{P90 Latency normalized to the SLOs.}
    \label{fig/evaluation/latency_reduction}
    \Description{None}
\end{figure}

\textbf{Summarization.}
We evaluated \sysname on the summarization task, with results presented in Figure~\ref{fig/evaluation/end-to-end/summarization_goodput}. \sysname improves goodput by 20–47\% over PD aggregation and by 30–77\% over PD disaggregation.

For SLO1, we used two P-heavy instances with a chunk size of 1024 and two with 256 to meet the 70 ms TPOT SLO. Compared to PD aggregation (chunk size 512), \sysname achieves 27\% and 35\% higher goodput for the 14B and 32B models, respectively (Figures~\ref{fig/evaluation/end-to-end/summarization_goodput/sub1},~\ref{fig/evaluation/end-to-end/summarization_goodput/sub2}), owing to reduced decode interference. Summarization’s long prompts require high prefill processing capacity, but in PD disaggregation, decode instances cannot process prefill, limiting capacity. As a result, \sysname outperforms PD disaggregation by 77\% and 74\% for the 14B and 32B models, respectively.
Under SLO2, targeting a stricter TPOT, we set the D-heavy chunk size to 128. In this scenario, \sysname delivers 26–47\% and 30–39\% higher goodput than PD aggregation and PD disaggregation, respectively (Figures~\ref{fig/evaluation/end-to-end/summarization_goodput/sub3},~\ref{fig/evaluation/end-to-end/summarization_goodput/sub4}).

\subsection{Latency Reduction}
\label{subsec/evaluation/latency_reduction}

Using latency-shifting scheduling policies, we optimize the latency of requests exceeding SLO constraints by selectively degrading the performance of suitable requests. Figures 17 and 18 show that \sysname reduces the 90th-percentile (P90) tail latency for TTFT and TPOT under maximum supported load, compared to PD aggregation and PD disaggregation.
Figure~\ref{fig/evaluation/ttft_reduction} shows \sysname achieves a $2.42\times$–$13.20\times$ reduction in TTFT relative to PD disaggregation, due to improved prefill processing capacity and the length-aware prefill scheduling algorithm, which degrades the degradable requests.
Figure~\ref{fig/evaluation/tpot_reduction} shows that \sysname reduces TPOT by $1.11\times$–$1.69\times$ compared to PD aggregation, due to low interference decode and precise flowing decode scheduling at the request level, which also avoids incorrect request degradation.

\begin{figure}[t!]
   \begin{minipage}{0.48\linewidth}
     \centering
     \includegraphics[width=0.98\textwidth]{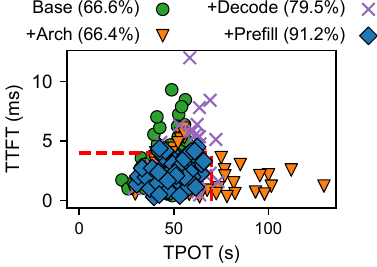}
     \caption{SLO attainment breakdown of proposed techniques: CP256 $\rightarrow$ +Arch $\rightarrow$ +Flowing Decode $\rightarrow$ +Length-Aware Prefill.}
     \label{fig/breakdown}
     \Description{None}
   \end{minipage}
   \hfill
   \begin{minipage}{0.48\linewidth}
     \centering
     \includegraphics[width=0.98\textwidth]{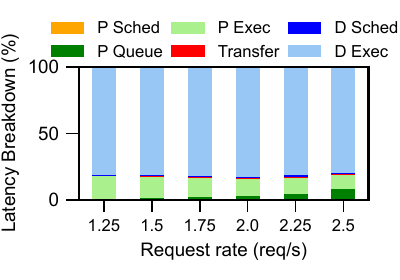}
     \caption{Latency breakdown of requests. Transfer overhead is negligible before the prefill queuing time blows up.}
     \label{fig/latency_breakdown}
     \Description{None}
   \end{minipage}
\end{figure}

\subsection{Performance Breakdown}
\label{subsec/evaluation/breakdown}

To demonstrate the effectiveness of our design, we perform a performance breakdown on the summarization task under SLO1 using Qwen2.5-14B. We start with a 4-instance PD aggregation configured with a chunk size of 256, and add our proposed techniques step by step. 
Figure~\ref{fig/breakdown} shows that we improved the SLO attainment rate from 66.6\% (Base) to 91.2\% by leveraging the support of the hybrid architecture and applying appropriate degradation decisions through the latency shifting scheduling policies.
The latency distribution for CP256 shows that using a small chunk size limits TPOT but causes unacceptable TTFT to exceed the SLO twice, due to limited prefill processing capacity.
With our hybrid architecture—setting chunk size to 1024 for the first two (P-heavy) instances—some requests have low TTFT but high TPOT (on P-heavy instances), while others show high TTFT and low TPOT (on D-heavy instances). Although this does not directly optimize latency, it provides a basis for further scheduling strategies.
Adding our flowing decode scheduling policy significantly reduces TPOT and increases SLO attainment by 12.9\%, as decode is prioritized on low-interference D-heavy instances and only select suitable requests are sent to P-heavy instances for TPOT degradation.
Finally, incorporating our length-aware prefill scheduling policy controls excessive TTFT and further raises SLO attainment by 11.7\%, by selectively degrading TTFT for suitable requests based on request length and instance queue status.

\subsection{Overhead Analysis}
\label{subsec/evaluation/overhead_analysis}

To demonstrate the low overhead of our proposed designs, we conducted a latency breakdown analysis for the SLO1 summarization task using Qwen2.5-14B. The primary sources of overhead are the KV cache transfer and the scheduling algorithm execution. As shown in Figure \ref{fig/latency_breakdown}, for each request, the transfer, prefill scheduling, and decode scheduling overheads are minimal, accounting for only 0.20\%, 0.01\%, and 0.89\% of the total request time, respectively. The low transfer overhead benefits from modern high-performance interconnects, while the low scheduling overhead is due to the lightweight nature of our algorithms.

\section{Related works}

\textbf{PD Aggregation.}
Orca~\cite{Orca} introduces continuous batching to improve throughput. FastServe~\cite{FastServe} employs iteration-level preemptive scheduling to reduce queuing delays for long-running tasks. NanoFlow~\cite{NanoFlow} decomposes batches into nano-batches to overlap computation, memory, and network usage, thereby enhancing GPU resource utilization. In contrast, \sysname focuses on optimizing goodput, which is orthogonal to these techniques. Sarathi-Serve~\cite{ChunkedPrefill} mitigates decode stalls in PD aggregation by introducing chunked prefill, which divides the prefill task into multiple chunks and piggybacks decode tasks with chunked prefill computation.
SOLA~\cite{SOLA} establishes an optimization model for PD aggregation, scheduling tasks based on the real-time status of requests and instances to balance TTFT and TPOT. 
Unlike these works that focus exclusively on PD aggregation, \sysname unifies PD aggregation, PD disaggregation, and a novel hybrid mode within a single architecture to achieve goodput-optimal performance under diverse SLO constraints.

\noindent\textbf{PD Disaggregation.}
DistServe~\cite{DistServe} and SplitWise~\cite{SplitWise} propose executing the prefill and decode phases on separate hardware resources to eliminate interference and enable phase-specific optimization. Adrenaline~\cite{liang2025adrenaline} optimizes the resource utilization of PD disaggregation via offloading the attention computation of decode to prefill instances. 
Unlike PD disaggregation, \sysname adopts a hybrid strategy that unifies PD aggregation and disaggregation to improve goodput by effectively balancing the trade-off between TTFT and TPOT.
Moreover, DynaServe~\cite{DynaServe}, which was developed concurrently with our system, introduces techniques to optimize the goodput of LLM service systems via balancing TTFT and TPOT.
It proposes splitting a request into two virtual sub-requests (e.g., the first containing prefill and a small part of early decode, and the second containing the remaining decode tasks), to balance token throughput and the \textit{time between token} (TBT) SLO constraint.
However, DynaServe assumes that the output length is known in advance, which is often unrealistic in real-world scenarios, as discussed in Challenge 2 (\S~\ref{subsec/challenges}). In contrast, our proposed \sysname dynamically controls per-request TPOT without requiring prior knowledge of output length, through an adaptive combination of multi-stage degradation flowing and optimization-triggered backflow.

\section{Conclusion}
\label{sec/conclusion}

We present a comprehensive study of PD aggregation and disaggregation in LLM serving, highlighting an inherent dilemma between these approaches for maximizing goodput under SLO constraints. To address it, we propose \sysname, a unified serving system that reallocates GPU resources to shift latency across different phases and requests, thereby increasing the SLO attainment rate. \sysname features a hybrid-mode inference, flowing decode scheduling, and length-aware prefill scheduling, enabling effective latency shifting for LLM serving. Experimental results demonstrate that \sysname improves goodput by up to 77\% compared to SOTA systems.

\bibliographystyle{ACM-Reference-Format}
\bibliography{sections/Reference}


\begin{thebibliography}{37}


\ifx \showCODEN    \undefined \def \showCODEN     #1{\unskip}     \fi
\ifx \showDOI      \undefined \def \showDOI       #1{#1}\fi
\ifx \showISBNx    \undefined \def \showISBNx     #1{\unskip}     \fi
\ifx \showISBNxiii \undefined \def \showISBNxiii  #1{\unskip}     \fi
\ifx \showISSN     \undefined \def \showISSN      #1{\unskip}     \fi
\ifx \showLCCN     \undefined \def \showLCCN      #1{\unskip}     \fi
\ifx \shownote     \undefined \def \shownote      #1{#1}          \fi
\ifx \showarticletitle \undefined \def \showarticletitle #1{#1}   \fi
\ifx \showURL      \undefined \def \showURL       {\relax}        \fi
\providecommand\bibfield[2]{#2}
\providecommand\bibinfo[2]{#2}
\providecommand\natexlab[1]{#1}
\providecommand\showeprint[2][]{arXiv:#2}

\bibitem[Abhyankar et~al\mbox{.}(2024)]%
        {Arxiv}
\bibfield{author}{\bibinfo{person}{Reyna Abhyankar}, \bibinfo{person}{Zijian He}, \bibinfo{person}{Vikranth Srivatsa}, \bibinfo{person}{Hao Zhang}, {and} \bibinfo{person}{Yiying Zhang}.} \bibinfo{year}{2024}\natexlab{}.
\newblock \bibinfo{title}{InferCept: Efficient Intercept Support for Augmented Large Language Model Inference}.
\newblock
\newblock
\showeprint[arxiv]{2402.01869}~[cs.LG]
\urldef\tempurl%
\url{https://arxiv.org/abs/2402.01869}
\showURL{%
\tempurl}


\bibitem[Agrawal et~al\mbox{.}(2024a)]%
        {Vidur}
\bibfield{author}{\bibinfo{person}{Amey Agrawal}, \bibinfo{person}{Nitin Kedia}, \bibinfo{person}{Jayashree Mohan}, \bibinfo{person}{Ashish Panwar}, \bibinfo{person}{Nipun Kwatra}, \bibinfo{person}{Bhargav~S. Gulavani}, \bibinfo{person}{Ramachandran Ramjee}, {and} \bibinfo{person}{Alexey Tumanov}.} \bibinfo{year}{2024}\natexlab{a}.
\newblock \showarticletitle{VIDUR: A LARGE-SCALE SIMULATION FRAMEWORK FOR LLM INFERENCE}. In \bibinfo{booktitle}{\emph{Proceedings of Machine Learning and Systems}}, \bibfield{editor}{\bibinfo{person}{P.~Gibbons}, \bibinfo{person}{G.~Pekhimenko}, {and} \bibinfo{person}{C.~De Sa}} (Eds.), Vol.~\bibinfo{volume}{6}. \bibinfo{pages}{351--366}.
\newblock
\urldef\tempurl%
\url{https://proceedings.mlsys.org/paper_files/paper/2024/file/b74a8de47d2b3c928360e0a011f48351-Paper-Conference.pdf}
\showURL{%
\tempurl}


\bibitem[Agrawal et~al\mbox{.}(2024b)]%
        {ChunkedPrefill}
\bibfield{author}{\bibinfo{person}{Amey Agrawal}, \bibinfo{person}{Nitin Kedia}, \bibinfo{person}{Ashish Panwar}, \bibinfo{person}{Jayashree Mohan}, \bibinfo{person}{Nipun Kwatra}, \bibinfo{person}{Bhargav~S. Gulavani}, \bibinfo{person}{Alexey Tumanov}, {and} \bibinfo{person}{Ramachandran Ramjee}.} \bibinfo{year}{2024}\natexlab{b}.
\newblock \showarticletitle{Taming throughput-latency tradeoff in LLM inference with sarathi-serve}. In \bibinfo{booktitle}{\emph{Proceedings of the 18th USENIX Conference on Operating Systems Design and Implementation}} (Santa Clara, CA, USA) \emph{(\bibinfo{series}{OSDI'24})}. \bibinfo{publisher}{USENIX Association}, \bibinfo{address}{USA}, Article \bibinfo{articleno}{7}, \bibinfo{numpages}{18}~pages.
\newblock
\showISBNx{978-1-939133-40-3}


\bibitem[AI(2023)]%
        {PersonalAssistant}
\bibfield{author}{\bibinfo{person}{Inflection AI}.} \bibinfo{year}{2023}\natexlab{}.
\newblock \bibinfo{booktitle}{\emph{Inflection-1 Technical Report}}.
\newblock \bibinfo{type}{{T}echnical {R}eport}. \bibinfo{institution}{Inflection AI}.
\newblock
\urldef\tempurl%
\url{https://inflection.ai/assets/Inflection-1.pdf}
\showURL{%
\tempurl}
\newblock
\shownote{Accessed: 2024-05-10}.


\bibitem[Ainslie et~al\mbox{.}(2023)]%
        {GQA}
\bibfield{author}{\bibinfo{person}{Joshua Ainslie}, \bibinfo{person}{James Lee-Thorp}, \bibinfo{person}{Michiel de Jong}, \bibinfo{person}{Yury Zemlyanskiy}, \bibinfo{person}{Federico Lebr'on}, {and} \bibinfo{person}{Sumit~K. Sanghai}.} \bibinfo{year}{2023}\natexlab{}.
\newblock \showarticletitle{GQA: Training Generalized Multi-Query Transformer Models from Multi-Head Checkpoints}.
\newblock \bibinfo{journal}{\emph{ArXiv}}  \bibinfo{volume}{abs/2305.13245} (\bibinfo{year}{2023}).
\newblock
\urldef\tempurl%
\url{https://api.semanticscholar.org/CorpusID:258833177}
\showURL{%
\tempurl}


\bibitem[Chen et~al\mbox{.}(2021)]%
        {CodeGen1}
\bibfield{author}{\bibinfo{person}{Mark Chen}, \bibinfo{person}{Jerry Tworek}, \bibinfo{person}{Heewoo Jun}, \bibinfo{person}{Qiming Yuan}, \bibinfo{person}{Henrique~Ponde de Oliveira~Pinto}, \bibinfo{person}{Jared Kaplan}, \bibinfo{person}{Harri Edwards}, \bibinfo{person}{Yuri Burda}, \bibinfo{person}{Nicholas Joseph}, \bibinfo{person}{Greg Brockman}, \bibinfo{person}{Alex Ray}, \bibinfo{person}{Raul Puri}, \bibinfo{person}{Gretchen Krueger}, \bibinfo{person}{Michael Petrov}, \bibinfo{person}{Heidy Khlaaf}, \bibinfo{person}{Girish Sastry}, \bibinfo{person}{Pamela Mishkin}, \bibinfo{person}{Brooke Chan}, \bibinfo{person}{Scott Gray}, \bibinfo{person}{Nick Ryder}, \bibinfo{person}{Mikhail Pavlov}, \bibinfo{person}{Alethea Power}, \bibinfo{person}{Lukasz Kaiser}, \bibinfo{person}{Mohammad Bavarian}, \bibinfo{person}{Clemens Winter}, \bibinfo{person}{Philippe Tillet}, \bibinfo{person}{Felipe~Petroski Such}, \bibinfo{person}{Dave Cummings}, \bibinfo{person}{Matthias Plappert}, \bibinfo{person}{Fotios Chantzis},
  \bibinfo{person}{Elizabeth Barnes}, \bibinfo{person}{Ariel Herbert-Voss}, \bibinfo{person}{William~Hebgen Guss}, \bibinfo{person}{Alex Nichol}, \bibinfo{person}{Alex Paino}, \bibinfo{person}{Nikolas Tezak}, \bibinfo{person}{Jie Tang}, \bibinfo{person}{Igor Babuschkin}, \bibinfo{person}{Suchir Balaji}, \bibinfo{person}{Shantanu Jain}, \bibinfo{person}{William Saunders}, \bibinfo{person}{Christopher Hesse}, \bibinfo{person}{Andrew~N. Carr}, \bibinfo{person}{Jan Leike}, \bibinfo{person}{Josh Achiam}, \bibinfo{person}{Vedant Misra}, \bibinfo{person}{Evan Morikawa}, \bibinfo{person}{Alec Radford}, \bibinfo{person}{Matthew Knight}, \bibinfo{person}{Miles Brundage}, \bibinfo{person}{Mira Murati}, \bibinfo{person}{Katie Mayer}, \bibinfo{person}{Peter Welinder}, \bibinfo{person}{Bob McGrew}, \bibinfo{person}{Dario Amodei}, \bibinfo{person}{Sam McCandlish}, \bibinfo{person}{Ilya Sutskever}, {and} \bibinfo{person}{Wojciech Zaremba}.} \bibinfo{year}{2021}\natexlab{}.
\newblock \bibinfo{title}{Evaluating Large Language Models Trained on Code}.
\newblock
\newblock
\showeprint[arxiv]{2107.03374}~[cs.LG]
\urldef\tempurl%
\url{https://arxiv.org/abs/2107.03374}
\showURL{%
\tempurl}


\bibitem[DeepSeek-AI et~al\mbox{.}(2024)]%
        {MLA}
\bibfield{author}{\bibinfo{person}{DeepSeek-AI}, \bibinfo{person}{Aixin Liu}, \bibinfo{person}{Bei Feng}, \bibinfo{person}{Bin Wang}, \bibinfo{person}{Bingxuan Wang}, \bibinfo{person}{Bo Liu}, \bibinfo{person}{Chenggang Zhao}, \bibinfo{person}{Chengqi Dengr}, \bibinfo{person}{Chong Ruan}, \bibinfo{person}{Damai Dai}, \bibinfo{person}{Daya Guo}, \bibinfo{person}{Dejian Yang}, \bibinfo{person}{Deli Chen}, \bibinfo{person}{Dongjie Ji}, \bibinfo{person}{Erhang Li}, \bibinfo{person}{Fangyun Lin}, \bibinfo{person}{Fuli Luo}, \bibinfo{person}{Guangbo Hao}, \bibinfo{person}{Guanting Chen}, \bibinfo{person}{Guowei Li}, \bibinfo{person}{H. Zhang}, \bibinfo{person}{Hanwei Xu}, \bibinfo{person}{Hao Yang}, \bibinfo{person}{Haowei Zhang}, \bibinfo{person}{Honghui Ding}, \bibinfo{person}{Huajian Xin}, \bibinfo{person}{Huazuo Gao}, \bibinfo{person}{Hui Li}, \bibinfo{person}{Hui Qu}, \bibinfo{person}{J.~L. Cai}, \bibinfo{person}{Jian Liang}, \bibinfo{person}{Jianzhong Guo}, \bibinfo{person}{Jiaqi Ni}, \bibinfo{person}{Jiashi
  Li}, \bibinfo{person}{Jin Chen}, \bibinfo{person}{Jingyang Yuan}, \bibinfo{person}{Junjie Qiu}, \bibinfo{person}{Junxiao Song}, \bibinfo{person}{Kai Dong}, \bibinfo{person}{Kaige Gao}, \bibinfo{person}{Kang Guan}, \bibinfo{person}{Lean Wang}, \bibinfo{person}{Lecong Zhang}, \bibinfo{person}{Lei Xu}, \bibinfo{person}{Leyi Xia}, \bibinfo{person}{Liang Zhao}, \bibinfo{person}{Liyue Zhang}, \bibinfo{person}{Meng Li}, \bibinfo{person}{Miaojun Wang}, \bibinfo{person}{Mingchuan Zhang}, \bibinfo{person}{Minghua Zhang}, \bibinfo{person}{Minghui Tang}, \bibinfo{person}{Mingming Li}, \bibinfo{person}{Ning Tian}, \bibinfo{person}{Panpan Huang}, \bibinfo{person}{Peiyi Wang}, \bibinfo{person}{Peng Zhang}, \bibinfo{person}{Qihao Zhu}, \bibinfo{person}{Qinyu Chen}, \bibinfo{person}{Qiushi Du}, \bibinfo{person}{R.~J. Chen}, \bibinfo{person}{R.~L. Jin}, \bibinfo{person}{Ruiqi Ge}, \bibinfo{person}{Ruizhe Pan}, \bibinfo{person}{Runxin Xu}, \bibinfo{person}{Ruyi Chen}, \bibinfo{person}{S.~S. Li}, \bibinfo{person}{Shanghao Lu},
  \bibinfo{person}{Shangyan Zhou}, \bibinfo{person}{Shanhuang Chen}, \bibinfo{person}{Shaoqing Wu}, \bibinfo{person}{Shengfeng Ye}, \bibinfo{person}{Shirong Ma}, \bibinfo{person}{Shiyu Wang}, \bibinfo{person}{Shuang Zhou}, \bibinfo{person}{Shuiping Yu}, \bibinfo{person}{Shunfeng Zhou}, \bibinfo{person}{Size Zheng}, \bibinfo{person}{T. Wang}, \bibinfo{person}{Tian Pei}, \bibinfo{person}{Tian Yuan}, \bibinfo{person}{Tianyu Sun}, \bibinfo{person}{W.~L. Xiao}, \bibinfo{person}{Wangding Zeng}, \bibinfo{person}{Wei An}, \bibinfo{person}{Wen Liu}, \bibinfo{person}{Wenfeng Liang}, \bibinfo{person}{Wenjun Gao}, \bibinfo{person}{Wentao Zhang}, \bibinfo{person}{X.~Q. Li}, \bibinfo{person}{Xiangyue Jin}, \bibinfo{person}{Xianzu Wang}, \bibinfo{person}{Xiao Bi}, \bibinfo{person}{Xiaodong Liu}, \bibinfo{person}{Xiaohan Wang}, \bibinfo{person}{Xiaojin Shen}, \bibinfo{person}{Xiaokang Chen}, \bibinfo{person}{Xiaosha Chen}, \bibinfo{person}{Xiaotao Nie}, \bibinfo{person}{Xiaowen Sun}, \bibinfo{person}{Xiaoxiang Wang},
  \bibinfo{person}{Xin Liu}, \bibinfo{person}{Xin Xie}, \bibinfo{person}{Xingkai Yu}, \bibinfo{person}{Xinnan Song}, \bibinfo{person}{Xinyi Zhou}, \bibinfo{person}{Xinyu Yang}, \bibinfo{person}{Xuan Lu}, \bibinfo{person}{Xuecheng Su}, \bibinfo{person}{Y. Wu}, \bibinfo{person}{Y.~K. Li}, \bibinfo{person}{Y.~X. Wei}, \bibinfo{person}{Y.~X. Zhu}, \bibinfo{person}{Yanhong Xu}, \bibinfo{person}{Yanping Huang}, \bibinfo{person}{Yao Li}, \bibinfo{person}{Yao Zhao}, \bibinfo{person}{Yaofeng Sun}, \bibinfo{person}{Yaohui Li}, \bibinfo{person}{Yaohui Wang}, \bibinfo{person}{Yi Zheng}, \bibinfo{person}{Yichao Zhang}, \bibinfo{person}{Yiliang Xiong}, \bibinfo{person}{Yilong Zhao}, \bibinfo{person}{Ying He}, \bibinfo{person}{Ying Tang}, \bibinfo{person}{Yishi Piao}, \bibinfo{person}{Yixin Dong}, \bibinfo{person}{Yixuan Tan}, \bibinfo{person}{Yiyuan Liu}, \bibinfo{person}{Yongji Wang}, \bibinfo{person}{Yongqiang Guo}, \bibinfo{person}{Yuchen Zhu}, \bibinfo{person}{Yuduan Wang}, \bibinfo{person}{Yuheng Zou},
  \bibinfo{person}{Yukun Zha}, \bibinfo{person}{Yunxian Ma}, \bibinfo{person}{Yuting Yan}, \bibinfo{person}{Yuxiang You}, \bibinfo{person}{Yuxuan Liu}, \bibinfo{person}{Z.~Z. Ren}, \bibinfo{person}{Zehui Ren}, \bibinfo{person}{Zhangli Sha}, \bibinfo{person}{Zhe Fu}, \bibinfo{person}{Zhen Huang}, \bibinfo{person}{Zhen Zhang}, \bibinfo{person}{Zhenda Xie}, \bibinfo{person}{Zhewen Hao}, \bibinfo{person}{Zhihong Shao}, \bibinfo{person}{Zhiniu Wen}, \bibinfo{person}{Zhipeng Xu}, \bibinfo{person}{Zhongyu Zhang}, \bibinfo{person}{Zhuoshu Li}, \bibinfo{person}{Zihan Wang}, \bibinfo{person}{Zihui Gu}, \bibinfo{person}{Zilin Li}, {and} \bibinfo{person}{Ziwei Xie}.} \bibinfo{year}{2024}\natexlab{}.
\newblock \bibinfo{title}{DeepSeek-V2: A Strong, Economical, and Efficient Mixture-of-Experts Language Model}.
\newblock
\newblock
\showeprint[arxiv]{2405.04434}~[cs.CL]
\urldef\tempurl%
\url{https://arxiv.org/abs/2405.04434}
\showURL{%
\tempurl}


\bibitem[Gao et~al\mbox{.}(2024)]%
        {gao2024cachedattention}
\bibfield{author}{\bibinfo{person}{Bin Gao}, \bibinfo{person}{Zhuomin He}, \bibinfo{person}{Puru Sharma}, \bibinfo{person}{Qingxuan Kang}, \bibinfo{person}{Djordje Jevdjic}, \bibinfo{person}{Junbo Deng}, \bibinfo{person}{Xingkun Yang}, \bibinfo{person}{Zhou Yu}, {and} \bibinfo{person}{Pengfei Zuo}.} \bibinfo{year}{2024}\natexlab{}.
\newblock \showarticletitle{{Cost‑Efficient} Large Language Model Serving for Multi‑turn Conversations with {CachedAttention}}. In \bibinfo{booktitle}{\emph{2024 USENIX Annual Technical Conference (USENIX ATC ’24)}}. \bibinfo{address}{Santa Clara, CA}, \bibinfo{pages}{111--126}.
\newblock
\showISBNx{978-1-939133-41-0}


\bibitem[Hong et~al\mbox{.}(2025)]%
        {SOLA}
\bibfield{author}{\bibinfo{person}{Ke Hong}, \bibinfo{person}{Xiuhong Li}, \bibinfo{person}{Lufang Chen}, \bibinfo{person}{Qiuli Mao}, \bibinfo{person}{Guohao Dai}, \bibinfo{person}{Xuefei Ning}, \bibinfo{person}{Shengen Yan}, \bibinfo{person}{Yun Liang}, {and} \bibinfo{person}{Yu Wang}.} \bibinfo{year}{2025}\natexlab{}.
\newblock \showarticletitle{SOLA: Optimizing SLO Attainment for Large Language Model Serving with State-Aware Scheduling}. In \bibinfo{booktitle}{\emph{Eighth Conference on Machine Learning and Systems}}.
\newblock


\bibitem[Jin et~al\mbox{.}(2023)]%
        {OutputLengthPrediction3}
\bibfield{author}{\bibinfo{person}{Yunho Jin}, \bibinfo{person}{Chun-Feng Wu}, \bibinfo{person}{David Brooks}, {and} \bibinfo{person}{Gu-Yeon Wei}.} \bibinfo{year}{2023}\natexlab{}.
\newblock \showarticletitle{S3: increasing GPU utilization during generative inference for higher throughput}. In \bibinfo{booktitle}{\emph{Proceedings of the 37th International Conference on Neural Information Processing Systems}} (New Orleans, LA, USA) \emph{(\bibinfo{series}{NIPS '23})}. \bibinfo{publisher}{Curran Associates Inc.}, \bibinfo{address}{Red Hook, NY, USA}, Article \bibinfo{articleno}{791}, \bibinfo{numpages}{13}~pages.
\newblock


\bibitem[Kleidermacher and Zou(2025)]%
        {Translation2}
\bibfield{author}{\bibinfo{person}{Hannah~Calzi Kleidermacher} {and} \bibinfo{person}{James Zou}.} \bibinfo{year}{2025}\natexlab{}.
\newblock \bibinfo{title}{Science Across Languages: Assessing LLM Multilingual Translation of Scientific Papers}.
\newblock
\newblock
\showeprint[arxiv]{2502.17882}~[cs.AI]
\urldef\tempurl%
\url{https://arxiv.org/abs/2502.17882}
\showURL{%
\tempurl}


\bibitem[Kwon et~al\mbox{.}(2023)]%
        {vLLM}
\bibfield{author}{\bibinfo{person}{Woosuk Kwon}, \bibinfo{person}{Zhuohan Li}, \bibinfo{person}{Siyuan Zhuang}, \bibinfo{person}{Ying Sheng}, \bibinfo{person}{Lianmin Zheng}, \bibinfo{person}{Cody~Hao Yu}, \bibinfo{person}{Joseph Gonzalez}, \bibinfo{person}{Hao Zhang}, {and} \bibinfo{person}{Ion Stoica}.} \bibinfo{year}{2023}\natexlab{}.
\newblock \showarticletitle{Efficient Memory Management for Large Language Model Serving with PagedAttention}. In \bibinfo{booktitle}{\emph{Proceedings of the 29th Symposium on Operating Systems Principles}} (Koblenz, Germany) \emph{(\bibinfo{series}{SOSP '23})}. \bibinfo{publisher}{Association for Computing Machinery}, \bibinfo{address}{New York, NY, USA}, \bibinfo{pages}{611–626}.
\newblock
\showISBNx{9798400702297}
\urldef\tempurl%
\url{https://doi.org/10.1145/3600006.3613165}
\showDOI{\tempurl}


\bibitem[{LangChain}(2023)]%
        {Summarization}
\bibfield{author}{\bibinfo{person}{{LangChain}}.} \bibinfo{year}{2023}\natexlab{}.
\newblock \bibinfo{title}{LangChain Use Case: Summarization}.
\newblock \bibinfo{howpublished}{\url{https://www.langchain.com/use-cases/summarization}}.
\newblock
\newblock
\shownote{Accessed: 2023}.


\bibitem[Liang et~al\mbox{.}(2025)]%
        {liang2025adrenaline}
\bibfield{author}{\bibinfo{person}{Yunkai Liang}, \bibinfo{person}{Zhangyu Chen}, \bibinfo{person}{Pengfei Zuo}, \bibinfo{person}{Zhi Zhou}, \bibinfo{person}{Xu Chen}, {and} \bibinfo{person}{Zhou Yu}.} \bibinfo{year}{2025}\natexlab{}.
\newblock \showarticletitle{Injecting Adrenaline into LLM Serving: Boosting Resource Utilization and Throughput via Attention Disaggregation}.
\newblock \bibinfo{journal}{\emph{arXiv preprint arXiv:2503.20552}} (\bibinfo{year}{2025}).
\newblock
\urldef\tempurl%
\url{https://arxiv.org/abs/2503.20552}
\showURL{%
\tempurl}


\bibitem[Lin et~al\mbox{.}(2024)]%
        {DocumentAnalysis}
\bibfield{author}{\bibinfo{person}{Yiming Lin}, \bibinfo{person}{Madelon Hulsebos}, \bibinfo{person}{Ruiying Ma}, \bibinfo{person}{Shreya Shankar}, \bibinfo{person}{Sepanta Zeigham}, \bibinfo{person}{Aditya~G. Parameswaran}, {and} \bibinfo{person}{Eugene Wu}.} \bibinfo{year}{2024}\natexlab{}.
\newblock \bibinfo{title}{Towards Accurate and Efficient Document Analytics with Large Language Models}.
\newblock
\newblock
\showeprint[arxiv]{2405.04674}~[cs.DB]
\urldef\tempurl%
\url{https://arxiv.org/abs/2405.04674}
\showURL{%
\tempurl}


\bibitem[{NVIDIA Corporation}(2023)]%
        {NCCL}
\bibfield{author}{\bibinfo{person}{{NVIDIA Corporation}}.} \bibinfo{year}{2023}\natexlab{}.
\newblock \bibinfo{title}{NVIDIA Collective Communications Library (NCCL)}.
\newblock \bibinfo{howpublished}{\url{https://developer.nvidia.com/nccl}}.
\newblock
\newblock
\shownote{Accessed: 2025-04-17}.


\bibitem[OpenAI(2022)]%
        {ChatGPT}
\bibfield{author}{\bibinfo{person}{OpenAI}.} \bibinfo{year}{2022}\natexlab{}.
\newblock \bibinfo{title}{Introducing {ChatGPT}}.
\newblock \bibinfo{howpublished}{\url{https://openai.com/blog/chatgpt}}.
\newblock
\newblock
\shownote{Accessed: 2025-04-15}.


\bibitem[OpenAI(2023)]%
        {GPT4}
\bibfield{author}{\bibinfo{person}{OpenAI}.} \bibinfo{year}{2023}\natexlab{}.
\newblock \bibinfo{title}{GPT-4}.
\newblock \bibinfo{howpublished}{\url{https://openai.com/index/gpt-4/}}.
\newblock
\newblock
\shownote{Accessed: 2025-05-08}.


\bibitem[Patel et~al\mbox{.}(2024)]%
        {SplitWise}
\bibfield{author}{\bibinfo{person}{Pratyush Patel}, \bibinfo{person}{Esha Choukse}, \bibinfo{person}{Chaojie Zhang}, \bibinfo{person}{Aashaka Shah}, \bibinfo{person}{Íñigo Goiri}, \bibinfo{person}{Saeed Maleki}, {and} \bibinfo{person}{Ricardo Bianchini}.} \bibinfo{year}{2024}\natexlab{}.
\newblock \showarticletitle{Splitwise: Efficient Generative LLM Inference Using Phase Splitting}. In \bibinfo{booktitle}{\emph{2024 ACM/IEEE 51st Annual International Symposium on Computer Architecture (ISCA)}}. \bibinfo{pages}{118--132}.
\newblock
\urldef\tempurl%
\url{https://doi.org/10.1109/ISCA59077.2024.00019}
\showDOI{\tempurl}


\bibitem[Qin et~al\mbox{.}(2024)]%
        {Mooncake}
\bibfield{author}{\bibinfo{person}{Ruoyu Qin}, \bibinfo{person}{Zheming Li}, \bibinfo{person}{Weiran He}, \bibinfo{person}{Mingxing Zhang}, \bibinfo{person}{Yongwei Wu}, \bibinfo{person}{Weimin Zheng}, {and} \bibinfo{person}{Xinran Xu}.} \bibinfo{year}{2024}\natexlab{}.
\newblock \bibinfo{title}{Mooncake: A KVCache-centric Disaggregated Architecture for LLM Serving}.
\newblock
\newblock
\showeprint[arxiv]{2407.00079}~[cs.DC]
\urldef\tempurl%
\url{https://arxiv.org/abs/2407.00079}
\showURL{%
\tempurl}


\bibitem[Qiu et~al\mbox{.}(2024)]%
        {OutputLengthPrediction1}
\bibfield{author}{\bibinfo{person}{Haoran Qiu}, \bibinfo{person}{Weichao Mao}, \bibinfo{person}{Archit Patke}, \bibinfo{person}{Shengkun Cui}, \bibinfo{person}{Saurabh Jha}, \bibinfo{person}{Chen Wang}, \bibinfo{person}{Hubertus Franke}, \bibinfo{person}{Zbigniew~T. Kalbarczyk}, \bibinfo{person}{Tamer Başar}, {and} \bibinfo{person}{Ravishankar~K. Iyer}.} \bibinfo{year}{2024}\natexlab{}.
\newblock \bibinfo{title}{Efficient Interactive LLM Serving with Proxy Model-based Sequence Length Prediction}.
\newblock
\newblock
\showeprint[arxiv]{2404.08509}~[cs.DC]
\urldef\tempurl%
\url{https://arxiv.org/abs/2404.08509}
\showURL{%
\tempurl}


\bibitem[Rozière et~al\mbox{.}(2024a)]%
        {Translation1}
\bibfield{author}{\bibinfo{person}{Baptiste Rozière}, \bibinfo{person}{Jonas Gehring}, \bibinfo{person}{Fabian Gloeckle}, \bibinfo{person}{Sten Sootla}, \bibinfo{person}{Itai Gat}, \bibinfo{person}{Xiaoqing~Ellen Tan}, \bibinfo{person}{Yossi Adi}, \bibinfo{person}{Jingyu Liu}, \bibinfo{person}{Romain Sauvestre}, \bibinfo{person}{Tal Remez}, \bibinfo{person}{Jérémy Rapin}, \bibinfo{person}{Artyom Kozhevnikov}, \bibinfo{person}{Ivan Evtimov}, \bibinfo{person}{Joanna Bitton}, \bibinfo{person}{Manish Bhatt}, \bibinfo{person}{Cristian~Canton Ferrer}, \bibinfo{person}{Aaron Grattafiori}, \bibinfo{person}{Wenhan Xiong}, \bibinfo{person}{Alexandre Défossez}, \bibinfo{person}{Jade Copet}, \bibinfo{person}{Faisal Azhar}, \bibinfo{person}{Hugo Touvron}, \bibinfo{person}{Louis Martin}, \bibinfo{person}{Nicolas Usunier}, \bibinfo{person}{Thomas Scialom}, {and} \bibinfo{person}{Gabriel Synnaeve}.} \bibinfo{year}{2024}\natexlab{a}.
\newblock \bibinfo{title}{Code Llama: Open Foundation Models for Code}.
\newblock
\newblock
\showeprint[arxiv]{2308.12950}~[cs.CL]
\urldef\tempurl%
\url{https://arxiv.org/abs/2308.12950}
\showURL{%
\tempurl}


\bibitem[Rozière et~al\mbox{.}(2024b)]%
        {CodeGen2}
\bibfield{author}{\bibinfo{person}{Baptiste Rozière}, \bibinfo{person}{Jonas Gehring}, \bibinfo{person}{Fabian Gloeckle}, \bibinfo{person}{Sten Sootla}, \bibinfo{person}{Itai Gat}, \bibinfo{person}{Xiaoqing~Ellen Tan}, \bibinfo{person}{Yossi Adi}, \bibinfo{person}{Jingyu Liu}, \bibinfo{person}{Romain Sauvestre}, \bibinfo{person}{Tal Remez}, \bibinfo{person}{Jérémy Rapin}, \bibinfo{person}{Artyom Kozhevnikov}, \bibinfo{person}{Ivan Evtimov}, \bibinfo{person}{Joanna Bitton}, \bibinfo{person}{Manish Bhatt}, \bibinfo{person}{Cristian~Canton Ferrer}, \bibinfo{person}{Aaron Grattafiori}, \bibinfo{person}{Wenhan Xiong}, \bibinfo{person}{Alexandre Défossez}, \bibinfo{person}{Jade Copet}, \bibinfo{person}{Faisal Azhar}, \bibinfo{person}{Hugo Touvron}, \bibinfo{person}{Louis Martin}, \bibinfo{person}{Nicolas Usunier}, \bibinfo{person}{Thomas Scialom}, {and} \bibinfo{person}{Gabriel Synnaeve}.} \bibinfo{year}{2024}\natexlab{b}.
\newblock \bibinfo{title}{Code Llama: Open Foundation Models for Code}.
\newblock
\newblock
\showeprint[arxiv]{2308.12950}~[cs.CL]
\urldef\tempurl%
\url{https://arxiv.org/abs/2308.12950}
\showURL{%
\tempurl}


\bibitem[Ruan et~al\mbox{.}(2025)]%
        {DynaServe}
\bibfield{author}{\bibinfo{person}{Chaoyi Ruan}, \bibinfo{person}{Yinhe Chen}, \bibinfo{person}{Dongqi Tian}, \bibinfo{person}{Yandong Shi}, \bibinfo{person}{Yongji Wu}, \bibinfo{person}{Jialin Li}, {and} \bibinfo{person}{Cheng Li}.} \bibinfo{year}{2025}\natexlab{}.
\newblock \bibinfo{title}{DynaServe: Unified and Elastic Tandem-Style Execution for Dynamic Disaggregated LLM Serving}.
\newblock
\newblock
\showeprint[arxiv]{2504.09285}~[cs.DC]
\urldef\tempurl%
\url{https://arxiv.org/abs/2504.09285}
\showURL{%
\tempurl}


\bibitem[ShareGPT(2023)]%
        {ShareGPT}
\bibfield{author}{\bibinfo{person}{ShareGPT}.} \bibinfo{year}{2023}\natexlab{}.
\newblock \bibinfo{title}{ShareGPT Teams}.
\newblock \bibinfo{howpublished}{\url{https://sharegpt.com/}}.
\newblock
\newblock
\shownote{Accessed: [Insert Access Date]}.


\bibitem[Strati et~al\mbox{.}(2024)]%
        {Dejavu}
\bibfield{author}{\bibinfo{person}{Foteini Strati}, \bibinfo{person}{Sara McAllister}, \bibinfo{person}{Amar Phanishayee}, \bibinfo{person}{Jakub Tarnawski}, {and} \bibinfo{person}{Ana Klimovic}.} \bibinfo{year}{2024}\natexlab{}.
\newblock \showarticletitle{D\'{e}j\`{a}Vu: KV-cache streaming for fast, fault-tolerant generative LLM serving}. In \bibinfo{booktitle}{\emph{Proceedings of the 41st International Conference on Machine Learning}} (Vienna, Austria) \emph{(\bibinfo{series}{ICML'24})}. \bibinfo{publisher}{JMLR.org}, Article \bibinfo{articleno}{1902}, \bibinfo{numpages}{27}~pages.
\newblock


\bibitem[Touvron et~al\mbox{.}(2023)]%
        {Llama2}
\bibfield{author}{\bibinfo{person}{Hugo Touvron}, \bibinfo{person}{Louis Martin}, \bibinfo{person}{Kevin Stone}, \bibinfo{person}{Peter Albert}, \bibinfo{person}{Amjad Almahairi}, \bibinfo{person}{Yasmine Babaei}, \bibinfo{person}{Nikolay Bashlykov}, \bibinfo{person}{Soumya Batra}, \bibinfo{person}{Prajjwal Bhargava}, \bibinfo{person}{Shruti Bhosale}, \bibinfo{person}{Dan Bikel}, \bibinfo{person}{Lukas Blecher}, \bibinfo{person}{Cristian~Canton Ferrer}, \bibinfo{person}{Moya Chen}, \bibinfo{person}{Guillem Cucurull}, \bibinfo{person}{David Esiobu}, \bibinfo{person}{Jude Fernandes}, \bibinfo{person}{Jeremy Fu}, \bibinfo{person}{Wenyin Fu}, \bibinfo{person}{Brian Fuller}, \bibinfo{person}{Cynthia Gao}, \bibinfo{person}{Vedanuj Goswami}, \bibinfo{person}{Naman Goyal}, \bibinfo{person}{Anthony Hartshorn}, \bibinfo{person}{Saghar Hosseini}, \bibinfo{person}{Rui Hou}, \bibinfo{person}{Hakan Inan}, \bibinfo{person}{Marcin Kardas}, \bibinfo{person}{Viktor Kerkez}, \bibinfo{person}{Madian Khabsa},
  \bibinfo{person}{Isabel Kloumann}, \bibinfo{person}{Artem Korenev}, \bibinfo{person}{Punit~Singh Koura}, \bibinfo{person}{Marie-Anne Lachaux}, \bibinfo{person}{Thibaut Lavril}, \bibinfo{person}{Jenya Lee}, \bibinfo{person}{Diana Liskovich}, \bibinfo{person}{Yinghai Lu}, \bibinfo{person}{Yuning Mao}, \bibinfo{person}{Xavier Martinet}, \bibinfo{person}{Todor Mihaylov}, \bibinfo{person}{Pushkar Mishra}, \bibinfo{person}{Igor Molybog}, \bibinfo{person}{Yixin Nie}, \bibinfo{person}{Andrew Poulton}, \bibinfo{person}{Jeremy Reizenstein}, \bibinfo{person}{Rashi Rungta}, \bibinfo{person}{Kalyan Saladi}, \bibinfo{person}{Alan Schelten}, \bibinfo{person}{Ruan Silva}, \bibinfo{person}{Eric~Michael Smith}, \bibinfo{person}{Ranjan Subramanian}, \bibinfo{person}{Xiaoqing~Ellen Tan}, \bibinfo{person}{Binh Tang}, \bibinfo{person}{Ross Taylor}, \bibinfo{person}{Adina Williams}, \bibinfo{person}{Jian~Xiang Kuan}, \bibinfo{person}{Puxin Xu}, \bibinfo{person}{Zheng Yan}, \bibinfo{person}{Iliyan Zarov}, \bibinfo{person}{Yuchen
  Zhang}, \bibinfo{person}{Angela Fan}, \bibinfo{person}{Melanie Kambadur}, \bibinfo{person}{Sharan Narang}, \bibinfo{person}{Aurelien Rodriguez}, \bibinfo{person}{Robert Stojnic}, \bibinfo{person}{Sergey Edunov}, {and} \bibinfo{person}{Thomas Scialom}.} \bibinfo{year}{2023}\natexlab{}.
\newblock \bibinfo{title}{Llama 2: Open Foundation and Fine-Tuned Chat Models}.
\newblock
\newblock
\showeprint[arxiv]{2307.09288}~[cs.CL]
\urldef\tempurl%
\url{https://arxiv.org/abs/2307.09288}
\showURL{%
\tempurl}


\bibitem[vllm project(2023)]%
        {vllm_project}
\bibfield{author}{\bibinfo{person}{vllm project}.} \bibinfo{year}{2023}\natexlab{}.
\newblock \bibinfo{title}{vLLM: Easy, Fast, and Cheap LLM Serving for Everyone}.
\newblock \bibinfo{howpublished}{\url{https://github.com/vllm-project/vllm}}.
\newblock
\newblock
\shownote{Accessed: 2025-04-14}.


\bibitem[Wang et~al\mbox{.}(2024a)]%
        {WorkloadStudy}
\bibfield{author}{\bibinfo{person}{Yuxin Wang}, \bibinfo{person}{Yuhan Chen}, \bibinfo{person}{Zeyu Li}, \bibinfo{person}{Zhenheng Tang}, \bibinfo{person}{Rui Guo}, \bibinfo{person}{Xin Wang}, \bibinfo{person}{Qiang Wang}, \bibinfo{person}{Amelie~Chi Zhou}, {and} \bibinfo{person}{Xiaowen Chu}.} \bibinfo{year}{2024}\natexlab{a}.
\newblock \showarticletitle{Towards Efficient and Reliable LLM Serving: A Real-World Workload Study}.
\newblock \bibinfo{journal}{\emph{ArXiv}}  \bibinfo{volume}{abs/2401.17644} (\bibinfo{year}{2024}).
\newblock
\urldef\tempurl%
\url{https://api.semanticscholar.org/CorpusID:271710854}
\showURL{%
\tempurl}


\bibitem[Wang et~al\mbox{.}(2024b)]%
        {MetricSurvey}
\bibfield{author}{\bibinfo{person}{Zhibin Wang}, \bibinfo{person}{Shipeng Li}, \bibinfo{person}{Yuhang Zhou}, \bibinfo{person}{Xue Li}, \bibinfo{person}{Rong Gu}, \bibinfo{person}{Nguyen Cam-Tu}, \bibinfo{person}{Chen Tian}, {and} \bibinfo{person}{Sheng Zhong}.} \bibinfo{year}{2024}\natexlab{b}.
\newblock \bibinfo{title}{Revisiting SLO and Goodput Metrics in LLM Serving}.
\newblock
\newblock
\showeprint[arxiv]{2410.14257}~[cs.LG]
\urldef\tempurl%
\url{https://arxiv.org/abs/2410.14257}
\showURL{%
\tempurl}


\bibitem[Wu et~al\mbox{.}(2024)]%
        {FastServe}
\bibfield{author}{\bibinfo{person}{Bingyang Wu}, \bibinfo{person}{Yinmin Zhong}, \bibinfo{person}{Zili Zhang}, \bibinfo{person}{Shengyu Liu}, \bibinfo{person}{Fangyue Liu}, \bibinfo{person}{Yuanhang Sun}, \bibinfo{person}{Gang Huang}, \bibinfo{person}{Xuanzhe Liu}, {and} \bibinfo{person}{Xin Jin}.} \bibinfo{year}{2024}\natexlab{}.
\newblock \bibinfo{title}{Fast Distributed Inference Serving for Large Language Models}.
\newblock
\newblock
\showeprint[arxiv]{2305.05920}~[cs.LG]
\urldef\tempurl%
\url{https://arxiv.org/abs/2305.05920}
\showURL{%
\tempurl}


\bibitem[Yu et~al\mbox{.}(2022)]%
        {Orca}
\bibfield{author}{\bibinfo{person}{Gyeong-In Yu}, \bibinfo{person}{Joo~Seong Jeong}, \bibinfo{person}{Geon-Woo Kim}, \bibinfo{person}{Soojeong Kim}, {and} \bibinfo{person}{Byung-Gon Chun}.} \bibinfo{year}{2022}\natexlab{}.
\newblock \showarticletitle{Orca: A Distributed Serving System for {Transformer-Based} Generative Models}. In \bibinfo{booktitle}{\emph{16th USENIX Symposium on Operating Systems Design and Implementation (OSDI 22)}}. \bibinfo{publisher}{USENIX Association}, \bibinfo{address}{Carlsbad, CA}, \bibinfo{pages}{521--538}.
\newblock
\showISBNx{978-1-939133-28-1}
\urldef\tempurl%
\url{https://www.usenix.org/conference/osdi22/presentation/yu}
\showURL{%
\tempurl}


\bibitem[Zhang et~al\mbox{.}(2025)]%
        {Tempo2025}
\bibfield{author}{\bibinfo{person}{Wei Zhang}, \bibinfo{person}{Zhiyu Wu}, \bibinfo{person}{Yi Mu}, \bibinfo{person}{Banruo Liu}, \bibinfo{person}{Myungjin Lee}, {and} \bibinfo{person}{Fan Lai}.} \bibinfo{year}{2025}\natexlab{}.
\newblock \showarticletitle{Tempo: Application‑aware LLM Serving with Mixed SLO Requirements}.
\newblock \bibinfo{journal}{\emph{arXiv preprint arXiv:2504.20068}} (\bibinfo{year}{2025}).
\newblock


\bibitem[Zhao et~al\mbox{.}(2024)]%
        {WildChat}
\bibfield{author}{\bibinfo{person}{Wenting Zhao}, \bibinfo{person}{Xiang Ren}, \bibinfo{person}{Jack Hessel}, \bibinfo{person}{Claire Cardie}, \bibinfo{person}{Yejin Choi}, {and} \bibinfo{person}{Yuntian Deng}.} \bibinfo{year}{2024}\natexlab{}.
\newblock \bibinfo{title}{WildChat: 1M ChatGPT Interaction Logs in the Wild}.
\newblock
\newblock
\showeprint[arxiv]{2405.01470}~[cs.CL]
\urldef\tempurl%
\url{https://arxiv.org/abs/2405.01470}
\showURL{%
\tempurl}


\bibitem[Zheng et~al\mbox{.}(2023)]%
        {OutputLengthPrediction2}
\bibfield{author}{\bibinfo{person}{Zangwei Zheng}, \bibinfo{person}{Xiaozhe Ren}, \bibinfo{person}{Fuzhao Xue}, \bibinfo{person}{Yang Luo}, \bibinfo{person}{Xin Jiang}, {and} \bibinfo{person}{Yang You}.} \bibinfo{year}{2023}\natexlab{}.
\newblock \showarticletitle{Response Length Perception and Sequence Scheduling: An LLM-Empowered LLM Inference Pipeline}.
\newblock \bibinfo{journal}{\emph{ArXiv}}  \bibinfo{volume}{abs/2305.13144} (\bibinfo{year}{2023}).
\newblock
\urldef\tempurl%
\url{https://api.semanticscholar.org/CorpusID:258833168}
\showURL{%
\tempurl}


\bibitem[Zhong et~al\mbox{.}(2024)]%
        {DistServe}
\bibfield{author}{\bibinfo{person}{Yinmin Zhong}, \bibinfo{person}{Shengyu Liu}, \bibinfo{person}{Junda Chen}, \bibinfo{person}{Jianbo Hu}, \bibinfo{person}{Yibo Zhu}, \bibinfo{person}{Xuanzhe Liu}, \bibinfo{person}{Xin Jin}, {and} \bibinfo{person}{Hao Zhang}.} \bibinfo{year}{2024}\natexlab{}.
\newblock \showarticletitle{DistServe: disaggregating prefill and decoding for goodput-optimized large language model serving}. In \bibinfo{booktitle}{\emph{Proceedings of the 18th USENIX Conference on Operating Systems Design and Implementation}} (Santa Clara, CA, USA) \emph{(\bibinfo{series}{OSDI'24})}. \bibinfo{publisher}{USENIX Association}, \bibinfo{address}{USA}, Article \bibinfo{articleno}{11}, \bibinfo{numpages}{18}~pages.
\newblock
\showISBNx{978-1-939133-40-3}


\bibitem[Zhu et~al\mbox{.}(2025)]%
        {NanoFlow}
\bibfield{author}{\bibinfo{person}{Kan Zhu}, \bibinfo{person}{Yufei Gao}, \bibinfo{person}{Yilong Zhao}, \bibinfo{person}{Liangyu Zhao}, \bibinfo{person}{Gefei Zuo}, \bibinfo{person}{Yile Gu}, \bibinfo{person}{Dedong Xie}, \bibinfo{person}{Tian Tang}, \bibinfo{person}{Qinyu Xu}, \bibinfo{person}{Zihao Ye}, \bibinfo{person}{Keisuke Kamahori}, \bibinfo{person}{Chien-Yu Lin}, \bibinfo{person}{Ziren Wang}, \bibinfo{person}{Stephanie Wang}, \bibinfo{person}{Arvind Krishnamurthy}, {and} \bibinfo{person}{Baris Kasikci}.} \bibinfo{year}{2025}\natexlab{}.
\newblock \bibinfo{title}{NanoFlow: Towards Optimal Large Language Model Serving Throughput}.
\newblock
\newblock
\showeprint[arxiv]{2408.12757}~[cs.DC]
\urldef\tempurl%
\url{https://arxiv.org/abs/2408.12757}
\showURL{%
\tempurl}


\end{thebibliography}

\end{document}